\documentclass[journal]{IEEEtran}
\usepackage{latexsym,,amssymb,amsmath,graphicx,epsf,cite,bbm,float,subfig}
\usepackage{ifpdf}
\usepackage{epstopdf}

\newcommand{\defi}{\stackrel{\bigtriangleup}{=}}
\newcommand{\nn}{\nonumber}
\newcommand{\eps}{\mbox{$\epsilon$}}
\newcommand{\sign}{\mathrm{sign}}
\newcommand{\Tr}{\mathrm{Tr}}
\newcommand{\diag}{\mathrm{diag}}
\newcommand{\bvec}{\mathrm{bvec}}
\newcommand{\col}{\mathrm{col}}

\renewcommand{\vec}[1]{\mbox{$\mathbf{#1}$}}
\newcommand{\vect}[1]{\mbox{\boldmath${#1}$}}
\newcommand{\vw}{\vec{w}}
\newcommand{\vhw}{\vect{\hat{\phi}}}
\newcommand{\vtw}{\vec{\tilde{w}}}
\newcommand{\vwo}{\vec{w_o}}
\newcommand{\uvw}{\vec{\underline{w}_o}}
\newcommand{\uuvw}{\vec{\underline{\underline{w}}_o}}

\newcommand{\vd}{\vec{d}}
\newcommand{\vn}{\vec{n}}
\newcommand{\vh}{\vec{h}}
\newcommand{\vht}{\underline{\vec{h}}}
\newcommand{\vb}{\vec{b}}
\newcommand{\vq}{\vec{q}}
\newcommand{\uvq}{\underline{\underline{\vec{q}_t}}}
\newcommand{\vv}{\vec{v}}
\newcommand{\vPi}{\vect{\Pi}}
\newcommand{\vDelta}{\vect{\Delta}}
\newcommand{\vsig}{\vect{\sigma}}
\newcommand{\ve}{\vec{e}}
\newcommand{\veps}{\vect{\eps}}
\newcommand{\vep}{\underline{\vec{e}}}
\newcommand{\va}{\vec{a}}
\newcommand{\vta}{\vec{\tilde{a}}}
\newcommand{\vc}{\vec{c}}
\newcommand{\vC}{\vec{C}}
\newcommand{\vu}{\vec{u}}
\newcommand{\vU}{\vec{U}}

\newcommand{\vphi}{\vect{\phi}}
\newcommand{\vtphi}{\vect{\tilde{\phi}}}
\newcommand{\vtpsi}{\vect{\tilde{\psi}}}
\newcommand{\RM}{\mbox{$\mathbbm{R}^M$}}

\newcommand{\mG}{\vec{G}}
\newcommand{\mQ}{\vec{Q}}
\newcommand{\mGd}{\vec{G_D}}
\newcommand{\mGc}{\vec{G_C}}
\newcommand{\mM}{\vec{M}}
\newcommand{\mN}{\vec{N}}
\newcommand{\mSig}{\vect{\Sigma}}
\newcommand{\mGam}{\vect{\Gamma}}
\newcommand{\mGamD}{\vect{\Gamma_D}}
\newcommand{\mGamC}{\vect{\Gamma_C}}
\newcommand{\mtSig}{\vect{\tilde{\Sigma}}}
\newcommand{\mSigD}{\vect{\Sigma^D_4}}
\newcommand{\mSigC}{\vect{\Sigma^C_4}}
\newcommand{\mOm}{\vect{\Omega}_t}
\newcommand{\umOm}{\underline{\vect{\Omega}}_t}
\newcommand{\mLam}{\vect{\Lambda}}
\newcommand{\mLamu}{\vect{\Lambda_u}}
\newcommand{\mLamc}{\vect{\Lambda_c}}

\newcommand{\eye}{\vec{I}}
\newcommand{\cN}{{\cal N}}
\newcommand{\cC}{\vec{{\cal C}}}
\newcommand{\cA}{{\cal A}}
\newcommand{\cX}{\vec{X}}
\newcommand{\cXu}{\vec{X_u}}
\newcommand{\cXd}{\vec{X_d}}
\newcommand{\cF}{\vec{F}}
\newcommand{\ctX}{\vec{Z}}
\newcommand{\ctXu}{\vec{Z_u}}
\newcommand{\ctXd}{\vec{Z_d}}
\newcommand{\cY}{\vec{Y}}
\newcommand{\cD}{\vec{D}}

\AtBeginDocument{
\addtolength{\abovedisplayskip}{-0.4ex}
\addtolength{\abovedisplayshortskip}{-0.4ex}
\addtolength{\belowdisplayskip}{-0.4ex}
\addtolength{\belowdisplayshortskip}{-0.4ex}
\addtolength{\belowcaptionskip}{-3ex}
}

\begin{document}

\title{Compressive Diffusion Strategies Over Distributed Networks for Reduced Communication Load}
\author{Muhammed O. Sayin, Suleyman S. Kozat*,~{\em Senior Member, IEEE}
\thanks{This work is in part supported by Turkish Academy of Sciences Outstanding Faculty Award Programme.

Muhammed O. Sayin and Suleyman S. Kozat are with the Department of Electrical and Electronics Engineering, Bilkent University, Bilkent, Ankara 06800 Turkey, Tel: +90 (312) 290-2336, Fax: +90 (312) 290-1223 (e-mail: sayin@ee.bilkent.edu.tr, kozat@ee.bilkent.edu.tr).}}

\maketitle
\begin{abstract}
We study the compressive diffusion strategies over distributed networks based on the diffusion implementation and adaptive extraction of the information from the compressed diffusion data. We demonstrate that one can achieve a comparable performance with the full information exchange configurations, even if the diffused information is compressed into a scalar or a single bit. To this end, we provide a complete performance analysis for the compressive diffusion strategies. We analyze the transient, steady-state and tracking performance of the configurations in which the diffused data is compressed into a scalar or a single-bit. We propose a new adaptive combination method improving the convergence performance of the compressive diffusion strategies further. In the new method, we introduce one more freedom-of-dimension in the combination matrix and adapt it by using the conventional mixture approach in order to enhance the convergence performance for any possible combination rule used for the full diffusion configuration. We demonstrate that our theoretical analysis closely follow the ensemble averaged results in our simulations. We provide numerical examples showing the improved convergence performance with the new adaptive combination method.
\end{abstract}

\begin{IEEEkeywords}
Compressed diffusion, distributed network, performance analysis.
\end{IEEEkeywords}
\begin{center}
\bfseries EDICS Category: ASP-ANAL, NET-DISP
\end{center}

\section{Introduction}
\IEEEPARstart{D}{istributed} network of nodes provides enhanced convergence performance for the applications such as source tracking, environment monitoring, and source localization \cite{sayed2013,li2002,akyildiz2002,estrin2001}. In such a network, each node encounters possibly a different statistical profile, which provides broadened perspective on the monitored phenomena. In general, we would reach the best estimate with access to all observation data across the whole network since the observation of each node carries valuable information \cite{sayed_book}. In the distributed adaptive estimation framework, we distribute the processing over the network and allow the information exchange among the nodes so that the parameter estimate of each node converges to the best estimate \cite{estrin2001,lopes2008}.

In the distributed architectures, there are several approaches regulating the information exchange, e.g., diffusion implementation. The diffusion implementation defines a communication protocol in which only the nodes from a predefined neighborhood could exchange information with each other \cite{sayed2013,lopes2008,cattivelli2010,cattivelli2008,kalman,tu2012,zhao2012}. In this framework, each node performs a local adaptive filtering algorithm and improves its parameter estimation by fusing with the diffused parameter estimations of the neighboring nodes. The diffusion approach provides robustness against link failures and changing network topologies \cite{lopes2008}. However, the diffusion of the parameter vector within the neighborhoods results in high amount of communication load. For example, since each node diffuses information to the neighbors, the total average number of information exchange is given by $N\times\overline{n}$ where $\overline{n}$ is the average size of a neighborhood in a network of $N$ nodes \cite{sayin2013}.

We study the compressive diffusion strategies that achieve better trade-off in terms of the amount of cooperation and the required communication load \cite{sayin2013}. Unlike the full diffusion configuration, the compressed diffusion approach diffuses single-bit of information or a reduced dimensional data vector. The diffused data is generated through certain random projection of the local parameter estimation vector. Then, the neighboring nodes can adaptively construct the original parameter estimations based on the diffused information and fuse their individual estimates for the final estimate. This approach reduces the communication load in the spirit of the compressive sensing \cite{sayin2013,donoho}. The compression is lossy since we do not assume any sparseness or compressibility on the parameter estimation vector \cite{donoho,cevher2010}. However, the compressive diffusion approach achieves comparable convergence performance with the full diffusion configurations. Since the communication load increases far more in the large networks or highly connected network of nodes, the compressive diffusion strategies play a crucial role in achieving comparable convergence performance with significantly reduced communication load.

There exists several other approaches that reduce the communication load. In \cite{xie2013}, within a predefined neighborhood, the parameter estimate is quantized before the diffusion in order to avoid unlimited bandwidth requirement. In \cite{ribeiro}, authors transmit the sign of the innovation sequence in the decentralized estimation framework. In \cite{sayyadi}, in a consensus network, the relative difference between the states of the nodes is exchanged by using a single bit of information. As distinct from the mentioned works, the compressive diffusion strategies substantially compress the diffused information and extract the information from the compressed data adaptively \cite{sayin2013}.

In this paper, we provide a complete performance analysis for the compressive diffusion strategies, which demonstrates comparable convergence performance of the compressed diffusion to the full information exchange configuration. We note that studying the performance of distributed networks with compressive diffusion strategies is not straight-forward since adaptive extraction of information from the diffused data brings in an additional adaptation level. Moreover, it is rather challenging for the single-bit diffusion strategy due to the nonlinear compression. However, we analyze the transient, steady-state and tracking performance of the configurations in which the diffused data is compressed into a scalar or a single-bit. We also propose a new adaptive combination method improving the performance for any conventional combination rule. In the compressive diffusion framework, we fuse the local estimates with the adaptively extracted information from substantially compressed diffusion data. The extracted information carries relatively less information than the original data. Hence, we introduce the confidence parameter concept, which adds one more freedom-of-dimension in the combination matrix. The confidence parameter determines how much we are confident with the local parameter estimation. Through the adaptation of the confidence parameter, we observe enormous enhancement in the convergence performance of the compressive diffusion strategies even for relatively long filter length.

Our main contributions include: 1) for Gaussian regressors, we analyze the transient, steady-state and tracking performance of scalar and single-bit diffusion techniques; 2) We demonstrate that our theoretical analysis accurately models the simulated results; 3) We propose a new adaptive combination method for compressive diffusion strategies, which achieves better trade-off in terms of the transient and steady state performance; 4) We provide numerical examples showing the enhanced convergence performance with the new adaptive combination method in our simulations.

We organize the paper as follows. In Section II, we explain the distributed network and diffusion implementation. In Section III, we introduce the compressive diffusion strategy, i.e., reduced-dimension and single-bit diffusion. In Section IV, we provide a global recursion model for the deviation parameters to facilitate the performance analysis. For Gaussian regressors, we analyze the mean-square convergence performance of the scalar and single-bit diffusion strategies in Section V and VI, respectively. In Section VII and VIII we analyze the steady-state and tracking performance of the scalar and single-bit diffusion approaches. In Section IX, we introduce the confidence parameter and propose a new adaptive combination method, improving the convergence performance of the compressive diffusion strategies. In Section X, we provide numerical examples demonstrating the match of theoretical and simulated results, and enhanced convergence performance with the new adaptive combination technique. We conclude the paper in Section XI with several remarks.

{\bf Notation:} Bold lower (or upper) case letters denote the column vectors (or matrices). For a vector $\vec{a}$ (or matrix $\vec{A}$), $\vec{a}^T$ (or $\vec{A}^T$) is its ordinary transpose. $\|\cdot\|$ and $\|\cdot\|_{\vec{A}}$ denote the $L_2$ norm and the weighted $L_2$ norm with the matrix $\vec{A}$, respectively (provided that $\vec{A}$ is positive-definite). We work with real data for notational simplicity. For a random variable $x$ (or vector $\vec{x}$), $E[x]$ (or $E[\vec{x}]$) represents its expectation. Here, $\Tr(\vec{A})$ denotes the trace of the matrix $\vec{A}$. The operator $\col\{\cdot\}$ creates a column vector or a matrix in which the arguments of $\col\{\cdot\}$ locate one under the other. For a matrix argument, $\diag\{\cdot\}$ operator returns the diagonal of the matrix as a vector and for a vector argument, it creates a diagonal matrix whose diagonal is the vector. The operator $\otimes$ takes the Kronecker tensor product of two matrices.
\vspace{-0.1in}
\section{Distributed Network}
Consider a network of $N$ nodes where each node $i$ observes a true parameter\footnote{Although we assume a time invariant unknown system vector, we also provide the tracking performance analysis for certain non-stationary models later in the paper.} $\vwo \in \RM$ through a linear model
\[
d_{i,t} = \vwo^T\vu_{i,t} + v_{i,t},
\]
where $v_{i,t}$ denotes the temporally and spatially white noise. We assume that the regression vector $\vu_{i,t} \in \RM$ is spatially and temporally uncorrelated with the other regressors and the observation noise. If we know the whole temporal and spatial data overall network, we can obtain the parameter of interest $\vwo$ by minimizing the following global cost with respect to the parameter estimate $\vw$:
\begin{align}
J_{\mathrm{glob}}(\vw) = \sum_{i=1}^{N} E\left[(d_{i,t} - \vw^T\vu_{i,t})^2\right].\label{eq:globalcost}
\end{align}
The stochastic gradient update for~\eqref{eq:globalcost} leads to the global least-mean square (LMS) algorithm as
\begin{align}
\vw_{t+1} = \vw_{t} + \mu \sum_{i=1}^N \vu_{i,t} \left( d_{i,t} - \vu_{i,t}^T\vw_t\right),\label{eq:globalLMS}
\end{align}
where $\mu > 0$ is the step size \cite{cattivelli2010}.
Note that~\eqref{eq:globalLMS} brings in significant communication burden by gathering the information overall network in a central processing unit. Additionally, centralized approach is not robust against the link failures and the changing network statistics \cite{estrin2001,lopes2008}. On the other hand, in the diffusion implementation framework, we utilize a protocol in which each node $i$ can only exchange information with nodes from its neighborhood $\cN_i$ with the convention $i \in \cN_i$~\cite{lopes2008,cattivelli2010}. This protocol distributes the processing to the nodes and provides tracking ability for time-varying statistical profiles~\cite{lopes2008}.

\begin{figure}[t!]
\centering
\includegraphics[width=1.5in]{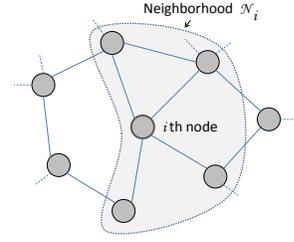}
\caption{Distributed network of nodes and the neighborhood ${\cal N}_i$}
\label{fig:network}
\end{figure}

Assuming the inner-node links are symmetric, we model the distributed network as an undirected graph where the nodes and the communication links correspond to its vertices and edges, respectively (See Fig.~\ref{fig:network}). In the distributed network, each node employs a local adaptation algorithm and benefits from the information diffused by the neighboring nodes in the construction of the final estimate \cite{lopes2008,cattivelli2010,cattivelli2008,kalman}. For example, in~\cite{lopes2008}, nodes \emph{diffuse their parameter estimate} to the neighboring nodes and each node $i$ performs the LMS algorithm given as
\begin{equation}
\vw_{i,t+1} = (\eye - \mu_i \vu_{i,t} \vu_{i,t}^T)\vphi_{i,t} + \mu_i d_{i,t}\vu_{i,t},\label{eq:DLMS}
\end{equation}
where $\mu_i > 0$ is the local step-size. The intermediate parameter vector $\vphi_{i,t}$ is generated through
\[
\vphi_{i,t} = \sum_{j \in \cN_i}\gamma_{i,j}\vw_{j,t}
\]
with $\gamma_{i,j}$'s are the combination weights such that $\sum_{j=1}^{N}\gamma_{i,j} = 1$ for all $i \in \{1,\cdots,N\}$. For a given network topology, the combination weights are determined according to certain combination rules such as uniform \cite{uniform}, the Metropolis \cite{xiao2004,metropolis}, relative-degree rules \cite{cattivelli2008} or adaptive combiners \cite{takahashi2010}.

We note that in \eqref{eq:DLMS} we could assign $\vphi_{i,t}$ as the final estimate in which we adapt the local estimate through the local observation data and then we fuse with the diffused estimates to generate the final estimate. In \cite{cattivelli2010}, authors examine these approaches as combine-than-adapt (CTA) and adapt-than-combine (ATC) diffusion strategies, respectively. In this paper, we study the ATC diffusion strategy, however, the theoretical results hold for both the ATC and CTA cases for certain parameter changes provided later in the paper.

We emphasize that the diffusion of the parameter estimation vector also brings in high amount of communication load. In the next section, we introduce the compressive diffusion strategies enabling the adaptive construction of the required information from the reduced dimension diffusion.
\vspace{-0.1in}
\section{Compressive Diffusion}
We seek to estimate the parameter of interest $\vwo$ through the \emph{reduced dimension information exchange} within the neighborhoods. In the compressed diffusion approach, unlike the full diffusion scheme, we diffuse a significantly reduced amount of information. The diffused information is generated by certain projection operator (a matrix $\cC_{t+1}$ or a vector $\vc_{t+1}$). Then, the neighboring nodes of $j$ generate an estimate $\va_{j,t+1}$ through the diffused information by using an adaptive estimation algorithm as explained later in the chapter \cite{sayin2013}. We point out that the diffused information might have far smaller dimensions than the parameter estimation vector, which can reduce the communication load significantly. The constructed estimates, i.e., $\va_{j,t+1}$'s are linearly combined with the local parameter estimate through certain combination rules, similar to the full diffusion configuration.

\begin{figure}[t!]
\centering
\includegraphics[width=3in]{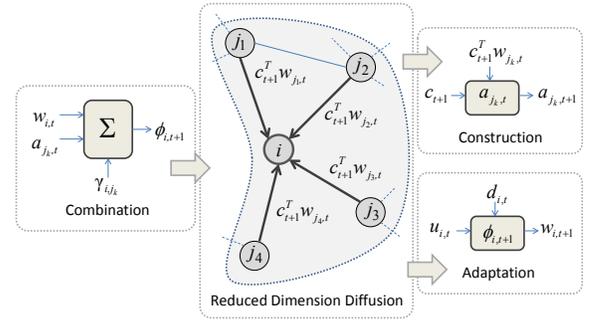}
\caption{CTA strategy in the compressive diffusion framework.}
\label{fig:strategy}
\end{figure}

\begin{figure}[t!]
\centering
\includegraphics[width=3in]{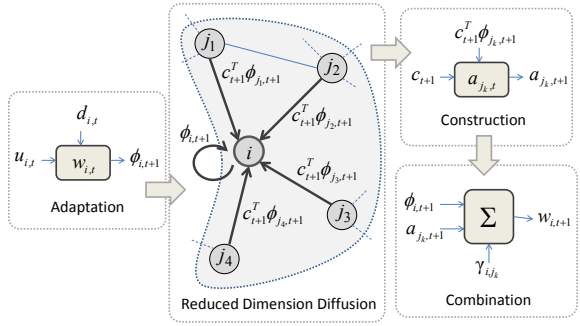}
\caption{ATC strategy in the compressive diffusion framework.}
\label{fig:strategy2}
\end{figure}

Different from the full diffusion configuration, in the new framework, nodes have access to the constructed estimates $\va_{j,t}$. Hence, in the compressive diffusion implementation, we update according to
\begin{align}
\vw_{i,t+1} = \arg\min_{\vw_i} &\left\{\gamma_{ii}\|\vw_i - \vw_{i,t}\|^2 + \sum_{j \in {\cal N}_i\setminus i} \gamma_{ij} \|\vw_{i} - \va_{j,t}\|^2 \right.\nn\\
&\left.+ \mu_i \left(d_{i,t} - \vu_{i,t}^T\vw_i\right)^2\right\}\label{adapt}
\end{align}
such that in the update we also minimize the Euclidean distance between the local parameter estimation $\vw_{i,t}$ and the constructed estimates $\va_{j,t}$ of the neighboring nodes. In order to simplify the optimization in \eqref{adapt}, we can replace the loss term $(d_{i,t} - \vu_{i,t}^T\vw_i)^2$ with the first order Taylor series expansion around $\va_{j,t}$, i.e.,
\begin{align}
(d_{i,t} - \vu_{i,t}^T\vw_i)^2 =& \bar{e}_{i,t}(\va_{j,t})^2 - 2 \bar{e}_{i,t}(\va_{j,t})\vu_{i,t}^T(\vw_i - \va_{j,t}) \nn\\
&+ O(\|\vw_{i}\|^2),\label{app1}
\end{align}
where we denote $\bar{e}_{i,t}(\va_{j,t}) \defi d_{i,t} - \vu_{i,t}^T\va_{j,t}$. Similarly, the first order Taylor series expansion around $\vw_{i,t}$ leads
\begin{equation}
(d_{i,t} - \vu_{i,t}^T\vw_i)^2 = e_{i,t}^2 - 2 e_{i,t}\vu_{i,t}^T(\vw_i - \vw_{i,t}) + O(\|\vw_{i}\|^2),\label{app2}
\end{equation}
where $e_{i,t} \defi d_{i,t} - \vu_{i,t}^T\vw_{i,t}$. Since $\sum_{j \in {\cal N}_i} \gamma_{ij} = 1$, the approximations \eqref{app1} and \eqref{app2} in \eqref{adapt} yields
\begin{align}
\vw_{i,t+1} =& \arg\min_{\vw_i} \left\{\gamma_{ii}\|\vw_i - \vw_{i,t}\|^2 + \sum_{j \in {\cal N}_i\setminus i} \gamma_{ij} \|\vw_{i} - \va_{j,t}\|^2 \right.\nn\\
&\hspace{-0.5cm}+ \mu_i \gamma_{ii}\left[e_{i,t}^2 - 2 e_{i,t}\vu_{i,t}^T(\vw_i-\vw_{i,t})\right]\nn\\
&\hspace{-0.5cm}\left.+ \mu_i\sum_{j \in {\cal N}_i\setminus i}\gamma_{ij}\left[\bar{e}_{i,t}(\va_{j,t})^2 - 2 \bar{e}_{i,t}(\va_{j,t})\vu_{i,t}^T(\vw_i-\va_{j,t})\right] \right\}.\label{adapt2}
\end{align}
The minimized term in \eqref{adapt2} is a convex function of $\vw_{i}$ and the Hessian matrix $2\eye_M \succ \vec{0}$ is positive definite. Hence, taking derivative and equating zero, we get the following update
\begin{align}
\vw_{i,t+1} =  \vphi_{i,t+1} + \mu_i\vu_{i,t}(d_{i,t} - \vu_{i,t}^T\vphi_{i,t+1}),\label{cta}
\end{align}
where
\begin{equation}
\vphi_{i,t+1} = \gamma_{ii}\vw_{i,t} + \sum_{j\in{\cal N}_i\setminus i}\gamma_{ij}\va_{j,t}, \label{cta2}
\end{equation}
which is similar to the distributed LMS algorithm \eqref{eq:DLMS}. Note that if we interchange $\vphi_{i,t}$ and $\vw_{i,t}$, in other words, when we assign the outcome of the combination as the final estimate rather than the outcome of the adaptation, we have the following algorithm:
\begin{align}
\vphi_{i,t+1} &=  \vw_{i,t} + \mu_i\vu_{i,t}(d_{i,t} - \vu_{i,t}^T\vw_{i,t}),\label{atc}\\
\vw_{i,t+1} &= \gamma_{ii}\vphi_{i,t+1} + \sum_{j\in{\cal N}_i\setminus i}\gamma_{ij}\va_{j,t+1}.\label{atc2}
\end{align}
We point out that \eqref{cta} and \eqref{cta2} are the CTA diffusion strategy while \eqref{atc} and \eqref{atc2} are the ATC diffusion strategy. Fig. \ref{fig:strategy} and \ref{fig:strategy2} summarize the compressive diffusion strategy for the CTA and ATC strategies where $j_k \in {\cal N}_i$. We next introduce different approaches to generate the diffused information (which are used to construct $a_{j,t+1}$'s).


In the compressive diffusion approach, irrespective of the final estimate we always diffuse the linear transformation of the outcome of the adaptation, e.g., we diffuse $z_{t+1} = \cC_{t+1}^T\vw_{i,t}$ in the CTA strategy and $z_{t+1} = \cC_{t+1}^T\vphi_{i,t+1}$ in the ATC strategy. Since we aim to use the most current parameter estimate in the construction of $\va_{j,t+1}$'s (since the most current estimate intuitively contains more information \cite{sayed2009}). We update according to
\begin{equation}
\va_{j,t+1} = \arg \min_{\va_j} \left\{\|\va_{j}-\va_{j,t}\|^2 + \eta_j \|\vec{z}_{t+1} - \cC_{t+1}^T\va_{j}\|^2\right\},\nn
\end{equation}
where we choose the diffused data as the desired signal and try to minimize the mean-square of the difference between the estimate $\vec{\hat{z}}_{t+1} = \cC_{t+1}^T\va_{j}$ and $\vec{z}_{t+1}$. The first order Taylor series approximation of the loss term $\|\vec{z}_{t+1}-\vec{\hat{z}}_{t+1}\|^2$ around $\va_{j,t}$ yields the following update
\begin{align}\label{eq:LMS}
\va_{j,t+1} = \va_{j,t} + \eta_j \cC_{t+1}\left(\vec{z}_{t+1} - \cC_{t+1}\va_{j,t}\right)
\end{align}
where $\eta_j > 0$ is the construction step size. We note that in \cite{sayin2013} the reduced dimension diffusion approach constructs $\va_{j,t+1}$'s through the minimum disturbance principle and resulted update involves $\left[\cC_{t+1}^T\cC_{t+1}\right]^{-1}$ as the normalization term. The constructed estimates $\va_{j,t+1}$'s are combined with the outcome of the local adaptation algorithm through \eqref{cta2} or \eqref{atc2}.

We next introduce a methods where the information exchange is only a single bit \cite{sayin2013}. When we construct $\va_{j,t}$ at node $j$, assuming $\va_{j,t}$'s are initialized with the same value, node $i \in {\cal N}_j$ has access to the exchanged estimate $\va_{j,t}$. Hence, we can perform the construction update at each neighboring node via the diffusion of the estimation error defined as
\begin{equation}\nn
\eps_{j,t+1}\defi z_{t+1} - \hat{z}_{t+1}.\footnote{In order to facilitate the performance analyzes, we redefine $\eps_{j,t+1}$ in \eqref{eq:eps2}.}
\end{equation}
Note that this does not influence the communication load, however, through the access to the exchange estimate $\va_{j,t+1}$ we can further reduce the communication load. Using the well-known sign algorithm \cite{sayed_book}, we can construct $\va_{j,t+1}$ as
\begin{align}\label{eq:sign}
\va_{j,t+1} = \va_{j,t} + \eta_{j}\vec{c}_{t+1}\mathrm{sign}(\eps_{j,t+1}),
\end{align}
where $\vc_{t+1}$ is the projection vector. Hence, we can repeat~\eqref{eq:sign} at each neighboring node via the diffusion of $z_{j,t+1} = \sign(\eps_{j,t+1})$ only and then we combine with the local estimate by using \eqref{cta2} or \eqref{atc2}.\\

\noindent
{\bf Remark 3.1:}
The compressive diffusion strategy reduces the communication load by constructing an estimate $\va_{i,t+1}$ corresponding to the original estimate $\vphi_{i,t+1}$ through the diffused information, i.e., the linear transformation of $\vphi_{i,t+1}$ with the projection operator $\cC_{t+1}$ or $\vc_{t+1}$. We note that the projection operator plays crucial role in the construction algorithms \eqref{eq:LMS} and \eqref{eq:sign}. We choose randomized projection operator that spans the whole parameter space in order to avoid biased convergence which degrades the performance \cite{sayin2013}. We point out that the randomized projection matrix (or vector) could be generated at each node synchronously provided that each node use the same \emph{seed} for the pseudo-random generator mechanism~\cite{barker}. Such seed exchanges and the synchronisation can be done periodically by using pilot signals without a serious increase in the communication load~\cite{pilot}.

In the next section, we introduce a global model gathering all network operations into a single update.
\vspace{-0.2in}
\section{Global Model}
For a vector projection operator, we write the reduced dimension~\eqref{eq:LMS} and single bit~\eqref{eq:sign} diffusion approaches for the ATC diffusion strategy in a compact form as
\begin{align}
\vphi_{i,t+1} &= \vw_{i,t} + \mu_i \vu_{i,t} e_{i,t}\label{eq:dist}\\
\va_{j,t+1}   &= \va_{j,t} + \eta_j \vc_{t+1} h(\eps_{j,t+1})\label{eq:construction}\\
\vw_{i,t+1}   &= \gamma_{i,i} \vphi_{i,t+1} + \sum_{j \in \cN_i\setminus i} \gamma_{i,j} \va_{j,t+1}\nn
\end{align}
where $e_{i,t} = d_{i,t} - \vu_{i,t}^{T}\vw_{i,t}$ and $\eps_{j,t+1} = \vc_{t+1}^T\left(\vphi_{j,t+1}-\va_{j,t}\right)$. For reduced dimension and single bit diffusion approaches, $h(\eps_{j,t+1}) = \eps_{j,t+1}$ and $h(\eps_{j,t+1}) = \sign(\eps_{j,t+1})$, respectively.

Next, we apply the following simplifications to facilitate the performance analyzes. First, we assume that at each node we use a different projection vector, e.g., for node j, we use $\vc_{j,t}$. Second, for sufficiently small $\mu_i$, we may substitute $\vphi_{i,t+1}$ with $\vphi_{i,t}$ in \eqref{eq:construction} (which is justified through simulations). With that simplifications, we can rewrite the update as
\begin{equation}\nn
\va_{j,t+1} = \va_{j,t} + \eta_j \vc_{j,t} h(\eps_{j,t}),
\end{equation}
where we redefine the construction error as
\begin{equation}\label{eq:eps2}
\eps_{j,t} \defi \vc_{j,t}^{T}(\vphi_{j,t}-\va_{j,t}).
\end{equation}
Note that we change $\vc_{j,t+1}$ with $\vc_{j,t}$ to be consistent with the introduced simplification.

For the state-space representation that collects all network operations into a single update, we define the following global parameters:
\begin{align*}
&\vphi_t = \col\{\vphi_{1,t},\ldots,\vphi_{N,t}\},\;\va_t = \col\{\va_{1,t},\ldots,\va_{N,t}\},\\
&\vw_t = \col\{\vw_{1,t},\ldots,\vw_{N,t}\},\;\uvw = \col\{\vwo, \ldots, \vwo\}
\end{align*}
with $MN\times 1$ dimensions and
\begin{align*}
&\ve_t = \col\{e_{1,t},\ldots,e_{N,t}\},\;\veps_t = \col\{\eps_{1,t},\ldots,\eps_{N,t}\},\\
&\vd_t = \col\{d_{1,t},\ldots,d_{N,t}\},\;\vv_t = \col\{v_{1,t},\ldots,v_{N,t}\}
\end{align*}
with $N\times 1$ dimensions. The combination matrix is given by
\begin{align*}
\mGam = \begin{bmatrix} \gamma_{11} & \cdots & \gamma_{1N}\\
                        \vdots      & \ddots & \vdots     \\
                        \gamma_{N1} & \cdots & \gamma_{NN}\end{bmatrix}
\end{align*}
and we denote $\mG \defi \mGam \otimes \eye_M$. Additionally, the regression and projection vectors yields the following $MN\times N$ global matrices
\begin{align*}
\vU_t \defi \begin{bmatrix} \vu_{1,t}&\cdots&\vec{0}\\ \vdots & \ddots & \vdots \\ \vec{0} & \cdots & \vu_{N,t}\end{bmatrix},\;
\vC_t \defi \begin{bmatrix} \vc_{1,t}&\cdots&\vec{0}\\ \vdots & \ddots & \vdots \\ \vec{0} & \cdots & \vc_{N,t}\end{bmatrix}.
\end{align*}
Indeed, we can model the network with compressive diffusion strategy as a larger network in which each node $i$ has an imaginary counterpart which diffuses $\va_{i,t}$ to the neighbors of $i$, which is similar to the full diffusion configuration. The real nodes only get information from the imaginary nodes and do not diffuse any information. In that case, the network can be modelled as a directed graph with asymmetric inner node links and the combination matrix is given by
\begin{align*}
\tilde{\mGam} = \begin{bmatrix} \mGamD & \mGamC \\ \vec{0} & \eye\end{bmatrix},
\end{align*}
where $\mGamD = \diag\{\mGam\}$ and $\mGamC = \mGam - \mGamD$. Then, we can write $\vw_t$ in terms of $\vphi_t$ and $\va_t$ as
\begin{equation}\label{eq:vw_t}
\vw_t = \mGd\vphi_t + \mGc\va_t,
\end{equation}
where $\mGd \defi \mGamD\otimes\eye_{M}$ and $\mGc \defi \mGamC \otimes\eye_M$. The state-space representation is given by
\begin{align*}
&\vphi_{t+1} = \vw_t + \mM \vU_t \ve_t,\\
&\va_{t+1} = \va_t + \mN \vC_t \vh(\veps_t),\\
&\vw_{t+1} = \mGd\vphi_{t+1} + \mGc\va_{t+1},
\end{align*}
where \[\mM \defi \diag\{[\mu_1,\ldots,\mu_N]\}\otimes\eye_M,\] \[\mN \defi \diag\{[\eta_1,\ldots,\eta_N]\}\otimes \eye_M\] and $\vh(\veps_t) = \col\{h(\eps_{1,t}),\cdots,h(\eps_{N,t})\}$. We obtain the global deviation vectors as
\begin{equation}\label{eq:dev}
\vtphi_t \defi \uvw - \vphi_t~\mbox{and}~\vta_t \defi \uvw - \va_t.
\end{equation}
Since $\mGam\underline{\vec{1}} = \underline{\vec{1}}$,
\begin{equation}\label{eq:Gw_o}
\mG\uvw = \uvw
\end{equation}
then the global deviation update yields
\begin{align}
\vtphi_{t+1} &= \mGd\vtphi_t + \mGc\vta_t - \mM\vU_t \ve_t,\label{eq:glob_est}\\
\vta_{t+1} &= \vta_t - \mN\vC_t\vh(\veps_t).\label{eq:glob_cons}
\end{align}

\begin{figure*}[!t]
\begin{align}
\overbrace{\begin{bmatrix} \vtphi_{t+1} \\ \vta_{t+1}\end{bmatrix}}^{\vtpsi_{t+1}} =
\overbrace{\begin{bmatrix} \mGd & \mGc \\ \vec{0} & \eye_{MN}\end{bmatrix}}^{\cX}
\overbrace{\begin{bmatrix} \vtphi_t \\ \vta_t\end{bmatrix}}^{\vtpsi_t} -
\overbrace{\begin{bmatrix} \mM & \vec{0} \\ \vec{0} & \mN\end{bmatrix}}^{\cD}
\overbrace{\begin{bmatrix} \vU_t & \vec{0} \\ \vec{0} & \vC_t\end{bmatrix}}^{\cY_t}
\overbrace{\begin{bmatrix} \ve_t \\ \vh(\veps_t)\end{bmatrix}}^{\vht(\ve_t,\veps_t)}
\label{equation}
\end{align}
\hrule
\end{figure*}

In \eqref{equation}, we represent the global deviation updates \eqref{eq:glob_est} and \eqref{eq:glob_cons} in a single equation or equivalently
\begin{equation}\label{eq:global}
\vtpsi_{t+1} = \cX\vtpsi_t - \cD\cY_t\vht(\ve_t,\veps_t),
\end{equation}
where $\vtpsi_t \defi \col\{\vtphi_t, \vta_t\}$. Based on the weighted-energy recursion of \eqref{eq:global}, in the next sections, we analyze the mean-square convergence performance of scalar and single-bit diffusion approaches separately for Gaussian regressors.
\vspace{-0.2in}
\section{Scalar Diffusion with Gaussian Regressors}
For the one-dimension diffusion approach, \eqref{eq:global} yields
\begin{equation}\label{eq:one-dim}
\vtpsi_{t+1} = \cX\vtpsi_t - \cD\cY_t\vep_t,
\end{equation}
where $\vep_t \defi \col\{\ve_t,\veps_t\}$. By \eqref{eq:vw_t}, \eqref{eq:dev} and \eqref{eq:Gw_o}, we note that $\ve_t$ is given by
\begin{align}
\ve_t = \vU_t^T(\mGd\vtphi_t + \mGc\vta_t) + \vv_t.\label{eq:e}
\end{align}
Similarly, we have
\begin{align}
\veps_t = \vC_t^T(-\vtphi_t + \vta_t).\label{eq:e2}
\end{align}
Hence, through \eqref{eq:e} and \eqref{eq:e2}, we obtain the global estimation error $\vep_t$ as
\begin{align}
\vep_t &= \begin{bmatrix} \vU_t & \vec{0} \\ \vec{0} & \vC_t \end{bmatrix}^T
         \underbrace{\begin{bmatrix} \mGd & \mGc \\ -\eye & \eye \end{bmatrix}}_{\ctX}
         \begin{bmatrix} \vtphi_t \\ \vta_t \end{bmatrix} +
         \underbrace{\begin{bmatrix} \vv_t \\ \vec{0}\end{bmatrix}}_{\vn_t} \nn\\
       &= \cY_t^T\ctX\vtpsi_t + \vn_t.\label{eq:e3}
\end{align}
Through \eqref{eq:e3}, we rewrite \eqref{eq:one-dim} as
\begin{align}\label{eq:rec1}
\vtpsi_{t+1} &= \cX\vtpsi_t - \cD\cY_t(\cY_t^T\ctX\vtpsi_t + \vn_t)\nn\\
             &= (\cX - \cD\cY_t\cY_t^T\ctX)\vtpsi_t - \cD\cY_t\vn_t.
\end{align}

We utilize the weighted-energy relation relating the energy of the error and deviation quantities in the performance analyzes through a weighting matrix $\mSig$. Then, we obtain
\begin{align*}
\vtpsi_{t+1}^T\mSig\vtpsi_{t+1} = &[(\cX - \cD\cY_t\cY_t^T\ctX)\vtpsi_t - \cD\cY_t\vn_t]^T\mSig\\
&\times[(\cX - \cD\cY_t\cY_t^T\ctX)\vtpsi_t - \cD\cY_t\vn_t]\\
= &\vtpsi_t^T(\cX - \cD\cY_t\cY_t^T\ctX)^T\mSig(\cX - \cD\cY_t\cY_t^T\ctX)\vtpsi_t\\
&-2\vn_t^T\cY_t^T\cD\mSig(\cX - \cD\cY_t\cY_t^T\ctX)\vtpsi_t\\
&+\vn_t^T\cY_t^T\cD\mSig\cD\cY_t\vn_t.
\end{align*}

Since we assume the observation noise $\vv_t$ is independent from the network statistics, the weighted energy relation for \eqref{eq:rec1} is given by
\begin{align}\label{eq:energy-rec}
E\|\vtpsi_{t+1}\|_{\mSig}^2 = E\|\vtpsi_t\|_{\mSig'}^2 + E[\vn_t^T\cY_t^T\cD\mSig\cD\cY_t\vn_t]
\end{align}
where
\begin{align*}
\mSig' \defi &(\cX - \cD\cY_t\cY_t^T\ctX)^T\mSig(\cX - \cD\cY_t\cY_t^T\ctX)\\
       = &\cX^T\mSig\cX - \ctX^T\cY_t\cY_t^T\cD\mSig\cX - \cX^T\mSig\cD\cY_t\cY_t^T\ctX\\
         &+ \ctX^T\cY_t\cY_t^T\cD\mSig\cD\cY_t\cY_t^T\ctX.
\end{align*}
Apart form the weighting matrix $\mSig$, $\mSig'$ is a random due to the data dependence. We assume the spatial and temporal independence of the regression data $\vu_{i,t}$ and $\vc_{j,t}$ so that $\cY_t$ is independent of $\vtpsi_t$. Through that assumption we can replace $\mSig'$ by its mean value, i.e., $\mSig' = E[\mSig']$ \cite{sayed_book,lopes2008}. Hence, the weighting matrix is given by
\begin{align}\label{eq:energy-rec2}
\mSig' = &\cX^T\mSig\cX - \ctX^T E\left[\cY_t\cY_t^T\right]\cD\mSig\cX - \cX^T\mSig\cD E\left[\cY_t\cY_t^T\right]\ctX\nn\\
         &+ \ctX^T\cD E\left[\cY_t\cY_t^T\mSig\cY_t\cY_t^T\right]\cD\ctX.
\end{align}
Note that in the last term of right hand side (RHS) of \eqref{eq:energy-rec2} we take $\cD$'s out of the expectation thanks to the block diagonal structure of $\cD$ and $\cY_t\cY_t^T$.

In order to calculate certain data moments in \eqref{eq:energy-rec} and \eqref{eq:energy-rec2}, we assume spatially and temporally i.i.d. Gaussian regression data such that
\begin{align}
\mLamu &\defi E[\vU_t\vU_t^T] = \diag\{[\sigma_{u,1}^2,\sigma_{u,2}^2,\ldots,\sigma_{u,N}^2]\}\otimes\eye_M\nn\\
\mLamc &\defi E[\vC_t\vC_t^T] = \diag\{[\sigma_{c,1}^2,\sigma_{c,2}^2,\ldots,\sigma_{c,N}^2]\}\otimes\eye_M.\nn
\end{align}
Then, we obtain
\begin{align*}
\mLam \defi E[\cY_t\cY_t^T] = \begin{bmatrix} \mLamu & \vec{0} \\ \vec{0} & \mLamc \end{bmatrix}.
\end{align*}

In the performance analysis, convenient vectorisation notation is used to exploit the diagonal structure of matrices \cite{sayed_book, tareqDataNorm}. In \eqref{eq:energy-rec}, \eqref{eq:energy-rec2}, matrices have block diagonal structures, thus, we use the block vectorisation operator $\bvec\{\cdot\}$ \cite{lopes2008}. Given an $NM\times NM$ block matrix
\begin{align*}
\mSig = \begin{bmatrix}\mSig_{11} & \ldots  & \mSig_{1N} \\
                       \vdots     & \ddots  & \vdots \\
                       \mSig_{N1} & \ldots  & \mSig_{NN}\end{bmatrix}
\end{align*}
where each block $\mSig_{ij}$ is a $M\times M$ block. Let $\vsig_{ij} = \mathrm{vec}\{\mSig_{ij}\}$ with standard $\mathrm{vec}\{\cdot\}$ operator and $\vsig_{j} = \col\{\vsig_{1j},\vsig_{2j},\ldots,\vsig_{Nj}\}$, then
\begin{align}\label{eq:bvec}
\bvec\{\mSig\} = \vsig = \col\{\vsig_1,\vsig_2,\ldots,\vsig_N\}.
\end{align}
We also use the \emph{block Kronecker product} of two block matrices $\vec{A}$ and $\vec{B}$, denoted by $\vec{A}\odot \vec{B}$. The $ij$-block is given by
\begin{align}
\left[\vec{A}\odot\vec{B}\right]_{ij} =\begin{bmatrix} \vec{A}_{ij}\otimes\vec{B}_{11} & \ldots & \vec{A}_{ij}\otimes\vec{B}_{1N}\\
\vdots & \ddots \vdots\\
 \vec{A}_{ij}\otimes\vec{B}_{N1} & \ldots & \vec{A}_{ij}\otimes\vec{B}_{NN} \end{bmatrix}.\label{eq:kron}
\end{align}
The block vectorisation operator $\bvec\{\cdot\}$ \eqref{eq:bvec} and the block Kronecker product~\eqref{eq:kron} are related by
\begin{align}
\bvec\{\vec{A}\mSig\vec{B}\} = (\vec{B}^T\odot \vec{A})\vsig\label{eq:bvec-kron}
\end{align}
and
\begin{align}\label{eq:trace}
\Tr\{\vec{A}^T\vec{B}\} = (\bvec\{\vec{A}\})^T\bvec\{\vec{B}\}.
\end{align}

The term in the RHS of \eqref{eq:energy-rec} yields
\begin{align*}
E\left[\vn_t^T\cY_t^T\cD\mSig\cD\cY_t\vn_t\right] = \Tr\left(\mLam\cD^2E\left[\vn_t\vn_t^T\right]\mSig\right)
\end{align*}
and let
\[
E\left[\vn_t\vn_t^T\right] = \vec{R_n} = \begin{bmatrix} \vec{R_v} & \vec{0} \\ \vec{0} & \vec{0}\end{bmatrix},
\]
where $\vec{R_v} \defi \diag\{\sigma_{v,1}^2,\ldots,\sigma_{v,N}^2\}\otimes\eye_M$. Then by \eqref{eq:trace},
\begin{align*}
E\left[\vn_t^T\cY_t^T\cD\mSig\cD\cY_t\vn_t\right] = \vb^T\vsig,
\end{align*}
where
\begin{equation}\label{eq:b}
\vb \defi \bvec\{\vec{R_n}\cD^2\mLam\}.
\end{equation}

The fourth-order moment in \eqref{eq:energy-rec2} yields
\[
\vec{A} = E\left[\cY_t\cY_t^T\mSig\cY_t\cY_t^T\right],
\]
where the $M\times M$ block is given by
\begin{align*}
[\vec{A}]_{ij} = \left\{\begin{array}{ll} 2\mLam_i\mSig_{ii}\mLam_i + \mLam_i\Tr\left(\mSig_{ii}\mLam_i\right) & i = j \\
                                               \mLam_i\mSig_{ij}\mLam_j & i \neq j \end{array}\right.
\end{align*}
thanks to the spatial and temporal independence of the regression data \cite{sayed_book}. We note that $\mLam$ could be denoted as $\mLam = \diag\{\mLam_1,\cdots,\mLam_N\}$ where $\mLam_i$ for $i=\{1,2,\ldots,N\}$ is $M\times M$ block matrix, e.g., $\mLam_1 = \sigma_{u,1}^2 \eye_M$. The $M\times M$ $ij$th block of $\mSig$ is denoted by $\mSig_{ij}$. Through \eqref{eq:kron}, \eqref{eq:trace}, we obtain
\begin{align*}
\bvec\{\vec{A}\} = \cA\vsig
\end{align*}
with $\cA = \diag\{\cA_1,\ldots,\cA_N\}$, $\cA_j = \diag\{\cA_{1j},\ldots,\cA_{Nj}\}$ and
\begin{align*}
\cA_{ij} = \left\{\begin{array}{ll} 2\mLam_i \otimes \mLam_i + \vec{\lambda}_i\vec{\lambda}_i^T & i = j \\
                                    \mLam_i \otimes \mLam_j & i \neq j \end{array}\right.
\end{align*}
where $\vec{\lambda}_i = \mathrm{vec}\{\mLam_i\}$.

Hence, the block vectorization of the weighting matrix $\mSig'$ \eqref{eq:energy-rec2} yields
\begin{align}
\bvec\{\mSig'\} = &\left(\cX^T\odot\cX^T - (\cX^T\odot\ctX^T)(\eye_{2MN}\odot\mLam\cD)\right.\nn\\
                  &-(\ctX^T\odot\cX^T)(\mLam\cD\odot\eye_{2MN}) \nn\\
                  &\left.+ (\ctX^T\odot\ctX^T)(\cD\odot\cD)\cA\right)\vsig.\nn
\end{align}
For notational simplicity, we change the weighted-norm notation such that $\|\vtphi_t\|_{\vsig}^2$ refers to $\|\vtphi_t\|_{\mSig}^2$ where $\vsig = \bvec\{\mSig\}$. As a result, we obtain the weighted-energy recursion as
\begin{align}
E\|\vtpsi_{t+1}\|_{\vsig}^2 &= E\|\vtpsi_t\|_{\cF\vsig}^2 + \vb^T\vsig\label{eq:energy1}\\
                        \cF &\defi \cX^T\odot\cX^T + (\ctX^T\odot\ctX^T)(\cD\odot\cD)\cA\nn\\
                            &-(\cX^T\odot\ctX^T)(\eye_{2MN}\odot\mLam\cD)\nn\\
                            &-(\ctX^T\odot\cX^T)(\mLam\cD\odot\eye_{2MN}).\label{eq:energy2}
\end{align}
Through \eqref{eq:energy1} and \eqref{eq:energy2}, we can analyze the learning, convergence and stability behavior of the network. Iterating the weighted-energy recursion, we obtain
\begin{align}
E\|\vtpsi_{t+1}\|_{\vsig}^2 &= E\|\vtpsi_t\|_{\cF\vsig}^2 + \vb^T\vsig \nn\\
E\|\vtpsi_{t}\|_{\cF\vsig}^2 &= E\|\vtpsi_{t-1}\|_{\cF^2\vsig}^2 + \vb^T\cF\vsig\nn\\
&\vdots\nn\\
E\|\vtpsi_{1}\|_{\cF^t\vsig}^2 &= E\|\vtpsi_0\|_{\cF^{t+1}\vsig}^2 + \vb^T\cF^t\vsig.\nn
\end{align}
Assuming the parameter estimates $\vphi_{i,t}$ and $\va_{i,t}$ are initialized with zeros, $E\|\vtpsi_0\|^2 = \|\uuvw\|^2$ where $\uuvw \defi \col\{\uvw,\uvw\}$. The iterations yield
\begin{equation}\label{eq:equation1}
E\|\vtpsi_{t+1}\|_{\vsig}^2 = \|\uuvw\|_{\cF^{t+1}\vsig}^2 + \vb^T\left(\sum_{k=0}^{t}\cF^k\right)\vsig.
\end{equation}
By \eqref{eq:equation1}, we reach the following final recursion:
\begin{equation}\label{eq:equation2}
E\|\vtpsi_{t+1}\|_{\vsig}^2 = E\|\vtpsi_{t}\|_{\vsig}^2 + \vb^T\cF^t\vsig - \|\uuvw\|_{\cF^t\left(\eye - \cF\right)\vsig}^2.
\end{equation}\\

\noindent
{\bf Remark 5.1:}
We note that \eqref{eq:equation2} is of essence since through the weighting matrix $\mSig$ we can extract information about the learning and convergence behavior of the network. In Table \ref{tab:sim2}, we tabulate the initial conditions (we assume the initial parameter vectors are set to $\vec{0}$) and the weighting matrices corresponding to various conventional performance measures.\\

\noindent
{\bf Remark 5.2:}
In this paper, \eqref{eq:equation2} provides a recursion for the weighted deviation parameter where we assign $\vphi_{i,t}$ as the final estimate instead of $\vw_{i,t}$, which implies the CTA strategy, however, the recursion also provides the performance of the ATC strategy with appropriate combination matrix $\mSig$ and the initial condition (See Table \ref{tab:sim2}).

\begin{table*}[t!]
\renewcommand{\arraystretch}{2}
\caption{Initial conditions and weighting matrices for different configurations.}
\label{tab:sim2}
\begin{center}
    \begin{tabular}{ | c || c | c | c || c | c | c |}
    \hline
    Framework & $E\|\vtpsi_t\|_{\mSig}^2$ &$E\|\vtpsi_0\|_{\mSig}^2$ & \mSig & $E\|\vtpsi_t\|_{\mSig}^2$ &$E\|\vtpsi_0\|_{\mSig}^2$ & \mSig\\ \hline
    CTA       & $ \frac{1}{N}E\|\vtphi_t\|^2$           &$\frac{1}{N}\|\uvw\|^2$            & $\frac{1}{N}\begin{bmatrix} \eye_{MN} & \vec{0} \\
                                                                                                           \vec{0}        & \vec{0} \end{bmatrix}$
              & $\frac{1}{N}E\|\vtphi_t\|^2_{\mLamu}$   &$\frac{1}{N}\|\uvw\|^2_{\mLamu}$   & $\frac{1}{N}\begin{bmatrix} \mLamu    & \vec{0} \\
                                                                                                           \vec{0}        & \vec{0} \end{bmatrix}$\\\hline
    ATC       & $\frac{1}{N}E\|\vtw_t\|^2$              &$\frac{1}{N}\|\uvw\|^2$            & $\frac{1}{N}\begin{bmatrix} \mGd^T\mGd     & \mGd^T\mGc \\
                                                                                                           \mGc^T\mGd     & \mGc^T\mGc \end{bmatrix}$
              & $\frac{1}{N}E\|\vtw_t\|^2_{\mLamu}$     &$\frac{1}{N}\|\uvw\|^2_{\mLamu}$   &$\frac{1}{N}\begin{bmatrix} \mGd^T\mLamu\mGd &
                                                                                                                        \mGd^T\mLamu\mGc \\
                                                                                                                        \mGc^T\mLamu\mGd & \mGc^T\mLamu\mGc \end{bmatrix}$\\\hline
    \end{tabular}
\end{center}
\end{table*}

Next, we analyze the mean-square convergence performance of the single-bit diffusion approach for Gaussian regressors.
\vspace{-0.1in}
\section{Single-Bit Diffusion with Gaussian Regressors}
The weighted-energy relation of \eqref{eq:global} yields
\begin{align}\label{eq:energy-rec3}
E\left[\vtpsi_{t+1}^T\mSig\vtpsi_{t+1}\right] &= E\left[\vtpsi_t^T\cX^T\mSig\cX\vtpsi_t\right]\nn\\
-& E\left[\vtpsi_t^T\cX^T\mSig\cD\cY_t\vht(\ve_t,\veps_t)\right]\nn\\
-& E\left[\vht^T(\ve_t,\veps_t)\cY_t^T\cD\mSig\cX\vtpsi_t\right]\nn \\
+& E\left[\vht^T(\ve_t,\veps_t)\cY_t^T\cD\mSig\cD\cY_t\vht(\ve_t,\veps_t)\right].
\end{align}
We evaluate RHS of \eqref{eq:energy-rec3} term by term in order to find the variance relation. Firstly, we partition the weighting matrix as follows:
\begin{align}\label{eq:partition}
\mSig = \begin{bmatrix}\mSig_1 & \mSig_2 \\ \mSig_3 & \mSig_4 \end{bmatrix}.
\end{align}
Through the partitioning \eqref{eq:partition}, we obtain
\begin{align}\label{eq:one}
E\left[\vtpsi_t^T\cX^T\mSig\cD\cY_t\vht(\ve_t,\veps_t)\right] &= E\left[\vtpsi_t^T\cXu^T\mSig_1\mM\vU_t\vU_t^T\ctXu\vtpsi_t\right]\nn\\
&\hspace{-2.3cm}+ E\left[\vtpsi_t^T\cXu^T\mSig_2\mN\vC_t\sign\left(\vC_t^T\ctXd\vtpsi_t\right)\right]\nn\\
&\hspace{-2.3cm}+ E\left[\vtpsi_t^T\cXd^T\mSig_3\mM\vU_t\vU_t^T\ctXu\vtpsi_t\right]\nn\\
&\hspace{-2.3cm}+ E\left[\vtpsi_t^T\cXd^T\mSig_4\mN\vC_t\sign\left(\vC_t^T\ctXd\vtpsi_t\right)\right],
\end{align}
where we partition $\cX$ such that $\cX = \col\{\cXu,\cXd\}$. We note that the second and fourth terms in the RHS of \eqref{eq:one} include the nonlinear $\sign(\cdot)$ function. It is not straight forward to evaluate the expectations with nonlinearity, thus, we introduce the following lemma.\\

\noindent
{\bf Lemma 1:}
{\em Under the assumption that step-sizes are sufficiently small and the filter is sufficiently long \cite{sayed_book}, the Price's theorem leads to
\begin{align}
&E\left[\vtpsi_t^T\cXu^T\mSig_2\mN\vC_t\sign\left(\vC_t^T\ctXd\vtpsi_t\right)\right] \nn\\
&\hspace{2cm}= E\left[\vtpsi_t^T\cXu^T\mSig_2\mN\mOm\vC_t\vC_t^T\ctXd\vtpsi_t\right], \label{eq:lem1}\\
&E\left[\vtpsi_t^T\cXd^T\mSig_4\mN\vC_t\sign\left(\vC_t^T\ctXd\vtpsi_t\right)\right] \nn\\
&\hspace{2cm}= E\left[\vtpsi_t^T\cXd^T\mSig_4\mN\mOm\vC_t\vC_t^T\ctXd\vtpsi_t\right], \label{eq:lem2}
\end{align}
where $\mOm$ is defined as
\begin{align} \nn
\mOm \defi \begin{bmatrix} \frac{E|\eps_{1,t}|}{E[\eps_{1,t}^2]}\eye_M & \cdots & \vec{0}_M \\
                           \vdots                                      & \ddots & \vdots    \\
                           \vec{0}_M                                   & \cdots & \frac{E|\eps_{N,t}|}{E[\eps_{N,t}^2]}\eye_M
                           \end{bmatrix}
\end{align}
}

{\em Proof:}
The proof is given in Appendix A. \hfill $\square$\\

By \eqref{eq:one}, \eqref{eq:lem1}, \eqref{eq:lem2}, the second term on the RHS of \eqref{eq:energy-rec3} is given by
\begin{align}\label{eq:one'}
E\left[\vtpsi_t^T\cX^T\mSig\cD\cY_t\vht(\ve_t,\veps_t)\right]& \nn\\
&\hspace{-1.5cm}= E\left[\vtpsi_t^T\cX^T\mSig\cD\umOm\cY_t\cY_t^T\ctX\vtpsi_t\right],
\end{align}
where $\umOm$ denotes
\begin{align*}
\umOm \defi \begin{bmatrix} \eye_{MN} & \vec{0} \\ \vec{0} & \mOm \end{bmatrix}.
\end{align*}
Similarly, the third term on the RHS of \eqref{eq:energy-rec3} is evaluated as
\begin{align}\label{eq:two'}
E\left[\vht^T(\ve_t,\veps_t)\cY_t^T\cD\mSig\cX\vtpsi_t\right]& \nn\\
&\hspace{-1.5cm}= E\left[\vtpsi_t^T\ctX^T\cY_t\cY_t^T\umOm\cD\mSig\cX\vtpsi_t\right].
\end{align}

Through partitioning, the last term on the RHS of \eqref{eq:energy-rec3} leads to
\begin{align}
E\left[\vht^T(\ve_t,\veps_t)\cY_t^T\cD\mSig\cD\cY_t\vht(\ve_t,\veps_t)\right]&\nn\\
&\hspace{-3.5cm}= E\left[\ve_t^T\vU_t^T\mM\mSig_1\mM\vU_t\ve_t\right] \nn\\
&\hspace{-3.2cm}+ E\left[\ve_t^T\vU_t^T\mM\mSig_2\mN\vC_t\sign(\veps_t)\right] \nn\\
&\hspace{-3.2cm}+ E\left[\sign(\veps_t)^T\vC_t^T\mN\mSig_3\mM\vU_t\ve_t\right] \nn\\
&\hspace{-3.2cm}+ E\left[\sign(\veps_t)^T\vC_t^T\mN\mSig_4\mN\vC_t\sign(\veps_t)\right].\nn
\end{align}\\

\noindent
{\bf Corollary 1:}
{\em Since $\vU_t$ and $\vC_t$ are independent from each other, similar to the Lemma 1, we obtain
\begin{align}\label{eq:three}
E\left[\vht^T(\ve_t,\veps_t)\cY_t^T\cD\mSig\cD\cY_t\vht(\ve_t,\veps_t)\right]&\nn\\
&\hspace{-3.5cm}= E\left[\ve_t^T\vU_t^T\mM\mSig_1\mM\vU_t\ve_t\right]\nn\\
&\hspace{-3.2cm}+ E\left[\ve_t^T\vU_t^T\mM\mSig_2\mN\mOm\vC_t\veps_t\right] \nn\\
&\hspace{-3.2cm}+ E\left[\veps_t^T\vC_t^T\mOm\mN\mSig_3\mM\vU_t\ve_t\right] \nn\\
&\hspace{-3.2cm}+ E\left[\sign(\veps_t)^T\vC_t^T\mN\mSig_4\mN\vC_t\sign(\veps_t)\right].
\end{align}
}

Because of the independence of the observation noise from the regression data, the first term on the RHS of \eqref{eq:three} yields
\begin{align}\label{eq:three-a}
E\left[\ve_t^T\vU_t^T\mM\mSig_1\mM\vU_t\ve_t\right] &= E\left[\vv_t^T\vU_t^T\mM\mSig_1\mM\vU_t\vv_t\right]\nn\\
&\hspace{-2.5cm}+ E\left[\vtpsi_t^T\ctXu^T\vU_t\vU_t^T\mM\mSig_1\mM\vU_t\vU_t^T\ctX_u\vtpsi_t\right].
\end{align}
For the last term on the RHS of \eqref{eq:three}, we introduce the following lemma.\\

\noindent
{\bf Lemma 2:}
{\em Through the Price's theorem, we obtain
\begin{align}\label{eq:lem3}
E\left[\sign(\veps_t)^T\vC_t^T\mN\mSig_4\mN\vC_t\sign(\veps_t)\right] &\nn\\
&\hspace{-3.7cm}= E\left[\vtpsi_t^T\ctXd^T\vC_t\vC_t^T\mN\mOm\mSigC\mOm\mN\vC_t\vC_t^T\ctXd\vtpsi_t\right]\nn\\
&\hspace{-3.3cm}+ E\left[\vec{1}^T\vC_t^T\mN\mSigD\mN\vC_t\vec{1}\right],
\end{align}
where $\mSigD$ is the block diagonal matrix of $\mSig_4$ such that
\begin{align}\nn
\mSigD = \begin{bmatrix} \vec{\Theta}_{11} & \cdots & \vec{0}_M \\
                         \vdots      & \ddots & \vdots \\
                         \vec{0}_M   & \cdots & \vec{\Theta}_{NN} \end{bmatrix}
\end{align}
with $\vec{\Theta}_{ii}$ is the $ii$'th $M\times M$ block of $\mSig_4$ and $\mSigC = \mSig_4 - \mSigD$.
}

{\em Proof:}
The proof is given in Appendix B. \hfill $\square$\\

As a result, by \eqref{eq:one'}, \eqref{eq:two'}, \eqref{eq:three}, \eqref{eq:three-a} and \eqref{eq:lem3}; \eqref{eq:energy-rec3} leads to
\begin{align}
E\|\vtpsi_{t+1}\|_{\mSig}^2 =& E\|\vtpsi_t\|_{\mSig'}^2 + E\left[\vv_t^T\vU_t^T\mM\mSig_1\mM\vU_t\vv_t\right]\nn\\
        &+E\left[\vec{1}^T\vC_t^T\mN\mSigD\mN\vC_t\vec{1}\right]\label{eq:energy-rec4a}
\end{align}
and
\begin{align*}
\mSig' =& \cX^T\mSig\cX - \cX^T\mSig\cD\umOm\cY_t\cY_t^T\ctX - \ctX^T\cY_t\cY_t^T\umOm\cD\mSig\cX\nn\\
        &+\ctX^T\cD\umOm\cY_t\cY_t^T\mtSig\cY_t\cY_t^T\umOm\cD\ctX,
\end{align*}
where $\mtSig$ denotes
\begin{align*}
\mtSig = \begin{bmatrix} \mSig_1 & \mSig_2 \\ \mSig_3 & \mSigC \end{bmatrix}.
\end{align*}

We again note that under the assumption that the regression data is spatially and temporally independent, we get $\mSig' = E[\mSig']$ which results
\begin{align}\label{eq:mSig}
\mSig' =& \cX^T\mSig\cX - \cX^T\mSig\cD\umOm\mLam\ctX - \ctX^T\mLam\umOm\cD\mSig\cX\nn\\
        &+\ctX^T\cD\umOm E\left[\cY_t\cY_t^T\mtSig\cY_t\cY_t^T\right]\umOm\cD\ctX
\end{align}
and denote $\vec{B} \defi E\left[\cY_t\cY_t^T\mtSig\cY_t\cY_t^T\right]$. Now, we resort to the vector notation, i.e., the block vectorisation operator $\bvec\{\cdot\}$ and the block Kronecker product. Hence, the block vectorization of the weighting matrix $\mSig'$ \eqref{eq:mSig} yields
\begin{align}
\bvec\{\mSig'\} &= \left(\cX^T\odot\cX^T - (\cX^T\odot\ctX^T)(\eye_{2MN}\odot\mLam\cD\umOm)\right.\nn\\
                  &\left.-(\ctX^T\odot\cX^T)(\mLam\cD\umOm\odot\eye_{2MN})\right)\vsig \nn\\
                  &+ (\ctX^T\odot\ctX^T)(\cD\odot\cD)(\umOm\odot\umOm)\bvec\{\vec{B}\}.\label{eq:weight2}
\end{align}
Block vectorisation of the matrix $\vec{B}$ is given by
\[
\bvec\{\vec{B}\}=\cA\,\bvec\{\mtSig\}.
\]
In order to denote $\bvec\{\mtSig\}$ in terms of $\vsig$, we introduce the following matrices:
\begin{align*}
\vec{K}_1 &\defi \col\{\vec{0}_{MN},\eye_{MN}\},\\
\vec{K}_2 &\defi \col\{\eye_{MN}, \vec{0}_{MN}\},\\
\vec{T_k} &\defi \diag\{\vec{0}_{(k-1)M},\eye_{M},\vec{0}_{(N-k)M}\}.
\end{align*}
We get $\mSigD$ and $\mtSig$ as
\begin{align}
\mSigD &= \sum_{k=1}^N \vec{T_k}\vec{K}_2^T\mSig\vec{K}_2\vec{T_k},\label{eq:1}\\
\mtSig &= \mSig - \vec{K}_2\mSigD\vec{K}_2^T.\label{eq:2}
\end{align}
By \eqref{eq:1} and \eqref{eq:2}, we obtain
\begin{align}
\bvec\{\mtSig\} &= \underbrace{\left(\eye - (\vec{K}_2\odot\vec{K}_2)\sum_{k=1}^N(\vec{T_k}\odot\vec{T_k})(\vec{K_2}^T\odot\vec{K}_2^T)\right)}_{\vec{K}}\vsig\nn\\
 &= \vec{K}\vsig.\label{eq:mSig'}
\end{align}

The $\vtpsi$-free terms in \eqref{eq:energy-rec4a} are evaluated as
\begin{align}
E\left[\vv_t^T\vU_t^T\mM\mSig_1\mM\vU_t\vv_t\right] &= \vb_1^T(\vec{K}_1^T\odot\vec{K}_1^T)\vsig,\label{eq:v}\\
E\left[\vec{1}^T\vC_t^T\mN\mSigD\mN\vC_t\vec{1}\right] &= \vb_2^T(\vec{K}_2^T\odot\vec{K}_2^T)\vsig,\label{eq:c}
\end{align}
where $\vb_1 \defi \bvec\{\vec{R_v}\mM^2\mLamu\}$ and $\vb_2 \defi \bvec\{\vec{1}\vec{1}^T\mN^2\mLamc\}$.

As a result, by \eqref{eq:weight2}, \eqref{eq:mSig'}, \eqref{eq:v} and \eqref{eq:c}, the weighted-energy relation is given by
\begin{align}
E\|\vtpsi_{t+1}\|_{\vsig}^2 =& E\|\vtpsi_t\|_{\cF_t\vsig}^2 + \vb^T\vsig \label{eq:recursion-a}\\
\cF_t = &\cX^T\odot\cX^T - (\cX^T\odot\ctX^T)(\eye_{2MN}\odot\mLam\cD\umOm)\nn\\
      &-(\ctX^T\odot\cX^T)(\mLam\cD\umOm\odot\eye_{2MN}) \nn\\
      &+(\ctX^T\odot\ctX^T)(\cD\odot\cD)(\umOm\odot\umOm)\cA\vec{K}\label{eq:recursion-b}\\
\vb = &(\vec{K}_1^T\odot\vec{K}_1^T)^T\vb_1 + (\vec{K}_2^T\odot\vec{K}_2^T)^T\vb_2.\label{eq:recursion-c}
\end{align}
Iterating the weighted-energy recursion \eqref{eq:recursion-a}, \eqref{eq:recursion-b} and \eqref{eq:recursion-c}, we obtain
\begin{align}
E\|\vtpsi_{t+1}\|_{\vsig}^2 &= E\|\vtpsi_t\|_{\cF_t\vsig}^2 + \vb^T\vsig \nn\\
E\|\vtpsi_{t}\|_{\cF_t\vsig}^2 &= E\|\vtpsi_{t-1}\|_{\cF_{t-1}\cF_t\vsig}^2 + \vb^T\cF_t\vsig\nn\\
&\vdots\nn\\
E\|\vtpsi_{1}\|_{\cF_1\ldots\cF_t\vsig}^2 &= E\|\vtpsi_0\|_{\cF_0\ldots\cF_t\vsig}^2 + \vb^T\cF_1\ldots\cF_t\vsig.\nn
\end{align}
In this part of the analyzes, we do not assume that the parameter vectors are initialized with zeros since such an assumption results in infinite terms in the $\mOm$ matrix. Hence, we initialize $\va_t$ with $\zeta\,\vec{1}_{MN\times 1}$ where $\zeta$ has a small value (See Table \ref{tab:sim1}).

The iterations yield
\begin{align}
E\|\vtpsi_{t+1}\|_{\vsig}^2 &= \|\vtpsi_0\|_{\vPi_t\vsig}^2 + \vb^T\vDelta_t\vsig,\label{eq:it1}\\
E\|\vtpsi_{t}\|_{\vsig}^2   &= \|\vtpsi_0\|_{\vPi_{t-1}\vsig}^2 + \vb^T\vDelta_{t-1}\vsig,\label{eq:it2}
\end{align}
where $\vPi_t \defi \prod_{i=0}^t \cF_i$ and $\vDelta_t \defi \eye + \cF_t + \cF_{t-1}\cF_t+\cdots+\cF_1\ldots\cF_t$. We note that $\vPi_t = \vPi_{t-1}\cF_t$ and $\vDelta_t = \vDelta_{t-1}\cF_t + \eye$. By \eqref{eq:it1} and \eqref{eq:it2}, we have the following recursion
\begin{align}
E\|\vtpsi_{t+1}\|_{\vsig}^2 =& E\|\vtpsi_t\|_{\vsig}^2 - \|\vtpsi_0\|_{\vPi_{t-1}(\eye-\cF_t)\vsig}^2\nn\\
&+\vb^T\left(\eye-\vDelta_{t-1}(\eye - \cF_t)\right)\vsig.\label{eq:equation3}
\end{align}
We point out that $\vPi_{-1} = \eye_{(2MN)^2}$ and $\vDelta_{-1} = \vec{0}_{(2MN)^2}$.\\

\begin{table}[t]
\renewcommand{\arraystretch}{2}
\caption{Initial conditions and weighting matrices for the performance measure of the construction update for the single-bit diffusion approach (for the scalar diffusion approach, set $\zeta = 0$) and the global MSD of the ATC diffusion strategy for the single-bit diffusion approach (for the scalar diffusion approach, see Table \ref{tab:sim2}).}
\label{tab:sim1}
\begin{center}
    \begin{tabular}{ | c | c | c |}
    \hline
    $E\|\vtpsi_t\|_{\mSig}^2$                   & $E\|\vtpsi_0\|_{\mSig}^2$     & \mSig \\\hline
    $\frac{1}{N}E\|\vta_t\|^2$                  & $\frac{1}{N}\|\uvw - \zeta \vec{1}\|^2$       & $\begin{bmatrix}  \vec{0}   & \vec{0} \\
    \vec{0}   & \frac{1}{N}\eye_{MN} \end{bmatrix}$ \\\hline
    $\sigma_{\veps_t}^2 = E[\veps_t^T\veps_t]$  & $\zeta\vec{1}^T\mLamc\vec{1}$ & $\begin{bmatrix} \mLamc & -\mLamc \\ -\mLamc & \mLamc \end{bmatrix}$ \\\hline
    $\frac{1}{N}E\|\vtw_t\|^2 $              &$\frac{1}{N}\|\uvw - \zeta\mGc\vec{1}\|^2$            & $\frac{1}{N}\begin{bmatrix} \mGd^T\mGd     & \mGd^T\mGc \\                                                                                                           \mGc^T\mGd     & \mGc^T\mGc \end{bmatrix}$\\\hline
\end{tabular}
\end{center}
\end{table}

\noindent
{\bf Remark 6.1:}
The iterations of \eqref{eq:equation3} requires the recalculation of $\cF_t$ for each time instants since $\cF_t$ changes with time because of $\umOm$ \eqref{eq:recursion-b}. Evaluating the expectations, $\mOm$ yields
\begin{align}\label{eq:mOm2}
\mOm =\sqrt{\frac{2}{\pi}}\begin{bmatrix} \frac{1}{\sigma_{\eps_1}} & \cdots & 0 \\
                                          \vdots                    & \ddots & \vdots    \\
                                          0                 & \cdots & \frac{1}{\sigma_{\eps_N}}
                           \end{bmatrix} \otimes \eye_M,
\end{align}
where $\sigma_{\eps_i}^2 = E[\eps_i^2]$. For analytical reasons, we approximate \eqref{eq:mOm2} as
\begin{align}\label{eq:mOm3}
\mOm \approx \sqrt{\frac{2}{\pi}}\frac{1}{(1/\sqrt{N})\sigma_{\veps_t}}\eye_{MN}
\end{align}
with
$\sigma_{\veps_t}^2 = E\left[\veps_t^T\veps_t\right] = E\|\vtpsi_t\|_{\vect{\xi}}^2 $
and
\begin{align*}
\vect{\xi} \defi \bvec\left\{\begin{bmatrix} \mLamc & -\mLamc \\ -\mLamc & \mLamc \end{bmatrix}\right\}.
\end{align*}
Hence, we can calculate $\cF_t$ by iterating the following
\begin{align}
E\|\vtpsi_{t+1}\|_{\vect{\xi}}^2 =& E\|\vtpsi_t\|_{\vect{\xi}}^2 - \|\vtpsi_0\|_{\vPi_{t-1}(\eye-\cF_t)\vect{\xi}}^2\nn\\
&+\vb^T\left(\eye-\vDelta_{t-1}(\eye - \cF_t)\right)\vect{\xi},\label{eq:equation4}
\end{align}
where $E\|\vtpsi_0\|_{\vect{\xi}}^2 = \zeta \vec{1}^T\mLamc\vec{1}$. In Table \ref{tab:sim1}, we tabulate the initial condition and the weighting matrix necessary for the recursion iterations \eqref{eq:equation4} of $\sigma_{\veps_t}^2 = E[\veps_t^T\veps_t]$.

\section{Steady-state Analysis}
At steady-state, \eqref{eq:energy1} yields
\begin{equation}\nn
E\|\vtpsi_{\infty}\|^2_{\left(\eye - \cF\right)\vsig} = \vb^T\vsig.
\end{equation}
In order to calculate the steady-state performance measure $E\|\vtpsi_{\infty}\|^2_{\vsig'}$ we choose the weighting matrix such that
\[
\vsig' = (\eye - \cF)\vsig
\]
then the steady-state performance measure is given by
\begin{equation}\label{eq:steady}
E\|\vtpsi_{\infty}\|^2_{\vsig'} = \vb^T(\eye - \cF)^{-1}\vsig'.
\end{equation}

Similar to \eqref{eq:steady}, the steady state mean square error $E[\veps_t^T\veps_t]$ for the single bit diffusion strategy is given by
\begin{equation}\label{eq:steady2}
E\|\vtpsi_{\infty}\|_{\vect{\xi}}^2 = \vb^T\left(\eye - \cF_{\infty}\right)^{-1}\vect{\xi}.
\end{equation}
We point out that $\cF_{\infty}$ depends on $E\|\vtpsi_{\infty}\|_{\vect{\xi}}^2$. Once we calculate $\cF_{\infty}$ numerically by \eqref{eq:steady2} or through rough approximations, we can obtain any steady state performance by \eqref{eq:steady}.

\section{Tracking Performance}
The diffusion implementation improves the ability of the network to track variations in the underlying statistical profiles \cite{lopes2008}. In this section, we analyze the tracking performance of the compressive diffusion strategies in a non-stationary environment. We assume a first-order random walk model, which is commonly used in the literature \cite{sayed_book}, for $\vwo(t)$ such that
\begin{equation}\nn
\vwo(t+1) = \vwo(t) + \vq_t,
\end{equation}
where $\vq_t \in \RM$ denotes a zero-mean vector process independent of the regression data and observation noise with covariance matrix $E[\vq_t\vq_t^T] = \mQ$. We introduce the global time-variant parameter vectors as $\uvw(t) = \col\{\vwo(t),\cdots,\vwo(t)\}$ and we have the global deviation vectors as $\vtphi_t = \uvw(t) - \vphi_t$ and $\vta_t = \uvw(t) - \va_t$. Then, by \eqref{eq:global}, we obtain
\begin{align}\label{eq:track}
\vtpsi_{t+1} = \cX\vtpsi_t -\cD\cY_t\vht(\ve_t,\veps_t) + \uvq,
\end{align}
where $\uvq \defi \col\{\vq_t,\cdots,\vq_t\}$ with $2MN\times 1$ dimensions. Since we assume that $\vq_t$ is independent from the regression data $\vu_{i,t}$, $\vc_{i,t}$ and the observation noise $\vv_{i,t}$ for all $i \in \{1,\cdots,N\}$, \eqref{eq:track} yields the following weighted-energy relation
\begin{align}\label{eq:track2}
E\left[\vtpsi_{t+1}^T\mSig\vtpsi_{t+1}\right] &= E\left[\vtpsi_t^T\cX^T\mSig\cX\vtpsi_t\right]\nn\\
-& E\left[\vtpsi_t^T\cX^T\mSig\cD\cY_t\vht(\ve_t,\veps_t)\right]\nn\\
-& E\left[\vht^T(\ve_t,\veps_t)\cY_t^T\cD\mSig\cX\vtpsi_t\right]\nn \\
+& E\left[\vht^T(\ve_t,\veps_t)\cY_t^T\cD\mSig\cD\cY_t\vht(\ve_t,\veps_t)\right]\nn\\
+& E\left[\uvq^T\mSig\uvq\right].
\end{align}
We note that \eqref{eq:track2} is similar to \eqref{eq:energy-rec3} except for the last term $E\left[\uvq^T\mSig\uvq\right]$. We denote $2N\times 2N$ matrix whose terms are $1$ as $\underline{\underline{\vec{1}}}_{2N} \defi [\vec{1},\cdots,\vec{1}]$. Then, the last term in \eqref{eq:track2} is given by $\vect{\rho}^T\vsig$ where $\vect{\rho} = \bvec\{\underline{\underline{\vec{1}}}_{2N}\otimes\mQ\}$. Through \eqref{eq:track2}, we get
\begin{equation}\label{eq:track3}
E\|\vtpsi_{t+1}\|_{\vsig}^2 = E\|\vtpsi_{t}\|_{\cF_t\vsig}^2 + \vb^T\vsig + \vect{\rho}^T\vsig.
\end{equation}
We define $\cF_t$ in \eqref{eq:energy2} and \eqref{eq:recursion-b} for scalar and single-bit diffusion strategies, respectively. Similarly, $\vb$ is introduced in \eqref{eq:b} and \eqref{eq:recursion-c} for the scalar (time-invariant) and single-bit diffusion strategies. We point out that \eqref{eq:track3} is different from \eqref{eq:energy1} and \eqref{eq:recursion-a} only for the term $\vect{\rho}^T\vsig$. As a result, at steady state, \eqref{eq:steady} and \eqref{eq:track3} leads
\begin{equation}\label{eq:track4}
E\|\vtpsi_{\infty}\|_{\vsig}^2 = (\vb + \vect{\rho})^T(\eye - \cF_{\infty})^{-1}\vsig.
\end{equation}
Through \eqref{eq:track4} and Table \ref{tab:sim2}, we can obtain the tracking performance of the network for the conventional performance measures.
We point out that in the full diffusion configuration, $\vect{\rho} = \bvec\{\underline{\underline{\vec{1}}}_{N}\otimes\mQ\}$.

In the next section, we introduce the confidence parameter and the adaptive combination method, which provides better trade-off in terms of transient and steady-state performance.
\vspace{-0.1in}
\section{Confidence Parameter and Adaptive Combination}
The cooperation among the nodes is not beneficial in general unless the cooperation rule is chosen properly \cite{sayed2013}. For example, uniform \cite{uniform}, the Metropolis \cite{xiao2004}, relative-degree rules \cite{cattivelli2008} and adaptive combiners \cite{takahashi2010} provide improved convergence performance relative to the no-cooperation configuration in which nodes aim to estimate the parameter of interest $\vwo$ without information exchange. However, the compressive diffusion strategies have a different diffusion protocol than the full diffusion configuration. At each node $i$, we combine the local estimates $\vphi_{i,t}$ with the constructed estimates $\va_{j,t}$ that track the local estimates $\vphi_{j,t}$ of the neighboring nodes, i.e., $j \in {\cal N}_i\setminus i$. Especially at the early stages of the adaptation, the constructed estimates carry far less information than the local estimates since they are not sufficiently close to the original estimates in the mean square sense. We point out that the global deviation equation of $\vphi_t$ could be written as
\begin{align}\label{eq:conf}
\vtphi_{t+1} = &\left(\eye - \mM\vU_t\vU_t^T\right)\mG\vtphi_t - \mM\vU_t\vv_t + \nn\\
&\left(\eye - \mM\vU_t\vU_t^T\right)\mGc\Delta\va_t,
\end{align}
where $\Delta\va_t \defi \vphi_t - \va_t$. In \eqref{eq:conf}, we observe that the compressive diffusion update includes one additional term, i.e., the last term on RHS of \eqref{eq:conf}, different from the full diffusion configuration. We can weaken the weight of the last term by arranging the combination matrix accordingly. Hence, we add one more freedom of dimension to the update by introducing a confidence parameter $\delta$. The confidence parameter determines the weight of the local estimates relative to the constructed estimates such that the new combination matrix $\mGam'$ is given by
\begin{equation}\label{eq:gam}
\mGam' = \delta \eye_N + (1-\delta)\mGam
\end{equation}
where $0\leq \delta \leq 1$. We note that $\delta = 1$, in which case we are confident with the local estimates, yields the no-cooperation scheme and $\delta = 0$ is the full diffusion configuration where we thrust the diffused information totally.

For the new combination matrix \eqref{eq:gam}, the combination of the local estimate and the constructed estimates \eqref{atc2} yields
\begin{align}\label{eq:comb}
\vw_{i,t+1} = &(1-\delta)\underbrace{\left[\gamma_{i,i}\vphi_{i,t+1}+\sum_{j\in{\cal N}_i\setminus i}\gamma_{i,j}\va_{j,t+1}\right]}_{\vhw_{i,t+1}}\nn \\
&+\delta \vphi_{i,t+1}
\end{align}
We note that \eqref{eq:comb} is a convex combination of the parameter vectors $\vhw_{i,t+1}$ and $\vphi_{i,t+1}$. Hence, we can adapt the convex combination weight $\delta$ using a stochastic gradient update \cite{garcia2005,garcia2006,silva2008,kozat}. Then, \eqref{eq:comb} yields
\begin{equation}\label{eq:comb2}
\vw_{i,t+1} = \delta_{i,t+1}\vphi_{i,t+1} + (1-\delta_{i,t+1})\vhw_{i,t+1}.
\end{equation}
In \cite{garcia2006}, authors update the combination weight indirectly through a sigmoidal function. Similarly, we re-parameterize the confidence parameter $\delta_{i,t}$ using the sigmoidal function \cite{sigmoid} and an unconstrained variable $\alpha_{i,t}$ such that
\begin{equation}\label{eq:sigmoid}
\delta_{i,t} = \frac{1}{1+e^{-\alpha_{i,t}}}.
\end{equation}
We train the unconstrained weight $\alpha_{i,t}$ using a stochastic gradient update minimizing $e_{i,t}^2 = \left(d_{i,t}-\vu_{i,t}^T\vw_{i,t}\right)^2$ as follows
\begin{align}\label{eq:update}
\alpha_{i,t+1} &= \alpha_{i,t} -\frac{1}{2}\mu_{\mathrm{cvx}}\frac{\partial e_{i,t}^2}{\partial \alpha_{i,t}}\nn\\
&= \alpha_{i,t} + \mu_{\mathrm{cvx}}e_{i,t}\vu_{i,t}^T(\vphi_{i,t}-\vhw_{i,t})\delta_{i,t}(1-\delta_{i,t}).
\end{align}
As a result, we combine the local and constructed estimates via \eqref{eq:comb2}, \eqref{eq:sigmoid} and \eqref{eq:update}.

In the next section, we provide numerical examples showing the match of the theoretical and simulated results, and the improved convergence performance with the adaptive confidence parameter.
\vspace{-0.2in}
\section{Numerical Examples}
In this section, we examine two distinct network scenarios where we demonstrate that the theoretical analysis accurately model the simulated results and confidence parameter provides significantly improved convergence performance. In the first example, we have a network of $5$ nodes where at each node $i$, we observe a stationary data $d_{i,t} = \vu_{i,t}^T\vwo + v_{i,t}$ for $i \in \{1,2,\cdots,N\}$. The regression data $\vu_{i,t}$ is zero-mean Gaussian with randomly chosen standard deviation $\sigma_{u_i}$, i.e., $\sigma_{u_i} = 0.1(\sqrt{10}-1){\cal U}[0,1] + 0.1$. The variance of the observation noise is $\sigma_{n_i}^2 = 10^{-3}$. In other words, the signal-to-noise ratio over the network varies around $10$ to $100$. The standard deviation of the projection operator is $\sigma_{c_i} = 1$. The parameter of interest $\vwo \in \mbox{$\mathbbm{R}^4$}$ is randomly chosen. Note that we examine a relatively small network with short filter length since the computational complexity of the theoretical performance relations \eqref{eq:equation2} and \eqref{eq:equation3} increases exponentially with the filter length $M$ and the network size $N$.

\begin{figure}[t!]
\centering
\includegraphics[width = 0.4\textwidth]{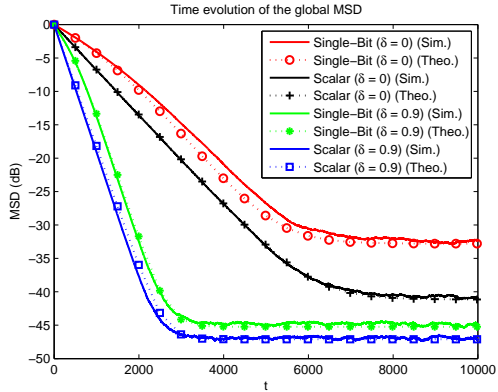}
\caption{Comparison of global MSD curves $1/N E\|\vtphi_t\|^2$ of the single-bit and scalar diffusion approaches for $\delta = 0$ and $\delta = 0.9$.}
\label{fig:sim8}
\end{figure}

In the no-cooperation configuration, the combination matrix is given by $\mGam_0 = \eye_N$. We use the Metropolis combination rule \cite{xiao2004} for the full diffusion configuration where the adjacency matrix of the network is given by
\[
\begin{bmatrix} 1&1&0&0&0\\	
                1&1&1&0&1\\
                0&1&1&1&0\\	
                0&0&1&1&0\\	
                0&1&0&0&1\end{bmatrix}.
\]
In the Metropolis rule \cite{metropolis}, the combination weights are chosen according to
\begin{align*}
\lambda_{i,j}=\left\{ \begin{array}{ll}
\frac{1}{\max\left\{n_i,n_j\right\}} & \mbox{if } j \in {\cal N}_i \setminus i, \\
0 & \mbox{if } j \notin {\cal N}_i,  \\
1-\sum_{j \in {\cal N}_i \setminus i} \lambda_{i,j} & \mbox{if } $i = j$,
\end{array}\right.
\end{align*}
where $n_i$ and $n_j$ denote the number of neighboring nodes for $i$ and $j$. For single-bit and one-dimension diffusion strategies we examine the convergence performance for the confidence parameter $\delta = 0$ and $\delta = 0.9$ in Fig. \ref{fig:sim8}. We choose the step sizes the same for the distributed LMS update \eqref{eq:dist} of all configurations at all nodes, i.e., $\mu_i = 0.042$. At each node, the step sizes for the construction update \eqref{eq:construction} are $\eta_i = 0.0015$ (for single-bit approach) and $\eta_i = 0.25$ (for one-dimension diffusion approach). For the single-bit diffusion approach, we set $\zeta = 0.001$ to initialize $\va_{j,t}$.

\begin{figure}[t!]
\centering
\includegraphics[width = 0.4\textwidth]{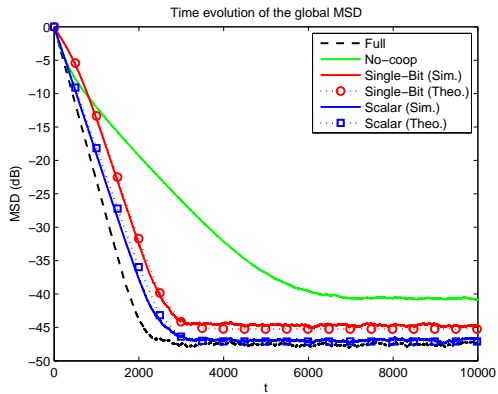}
\caption{Comparison of the global MSD curves of the no-cooperation, single-bit, scalar and full diffusion configurations in the CTA diffusion strategy.}
\label{fig:sim3-MSD}
\end{figure}

\begin{figure}[t!]
\centering
\includegraphics[width = 0.4\textwidth]{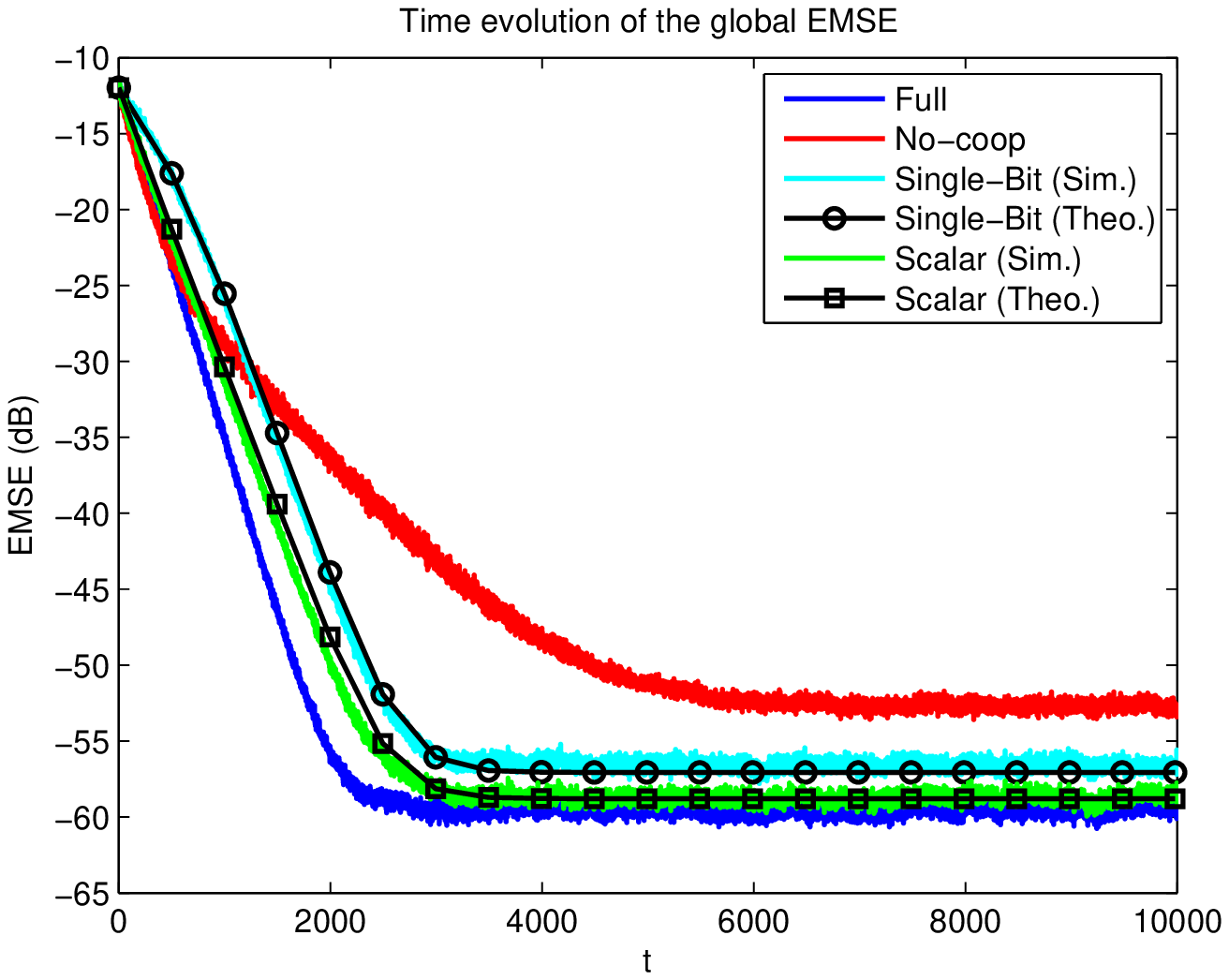}
\caption{Comparison of the global EMSE curves of the no-cooperation, single-bit, scalar and full diffusion configurations in the CTA diffusion strategy.}
\label{fig:sim3-EMSE}
\end{figure}

\begin{figure}[t!]
\centering
\includegraphics[width = 0.4\textwidth]{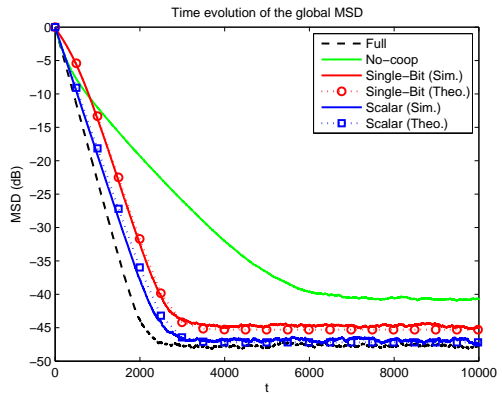}
\caption{Comparison of the global MSD curves of the no-cooperation, single-bit, scalar and full diffusion configurations in the ATC diffusion strategy.}
\label{fig:sim3-ATC}
\end{figure}

In Fig. \ref{fig:sim8}, we show the global MSD curves, i.e., $E\|\vtphi_t\|^2$, of the single-bit and scalar diffusion approaches and compare the performance for different $\delta$ values. The confidence parameter $\delta = 0.9$ implies that we give ten times more weight to the local estimate $\vphi_{i,t}$ than the constructed estimates $\va_{j,t}$ where $j \in {\cal N}_i \setminus i$. The Fig. \ref{fig:sim8} demonstrates that the confidence parameter $\delta = 0.9$ improves the convergence performance of the compressive diffusion strategies. For the same example, Fig. \ref{fig:sim3-MSD}, Fig. \ref{fig:sim3-EMSE} and Fig. \ref{fig:sim3-ATC} compare the convergence performance of single-bit and scalar diffusion strategies with the no-cooperation and full diffusion configurations for $\delta = 0.9$, which shows the match of the theoretical and ensemble averaged (we perform 200 independent trials) performance results. The Fig. \ref{fig:sim3-MSD} and Fig. \ref{fig:sim3-EMSE} show the time-evolution of the MSD and EMSE curves in the CTA diffusion strategy while the Fig. \ref{fig:sim3-ATC} displays the time-evolution of the MSD curves in the ATC diffusion strategy in which the theoretical curves \eqref{eq:equation2} and \eqref{eq:equation3} are iterated according to the Table \ref{tab:sim2} and \ref{tab:sim1}. We note that we obtain similar MSD curves in the CTA and ATC strategies. Since we set $\delta = 0.9$ and the outcomes of the adaptation and combination operations contain relatively close amount of information.

\begin{figure}[t!]
\centering
\includegraphics[width = 0.4\textwidth]{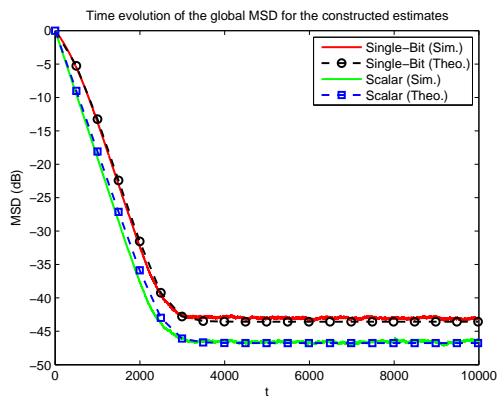}
\caption{The MSD curves of the construction estimate $1/N E\|\vta_t\|^2$ of the single-bit and scalar diffusion approaches.}
\label{fig:sim3-CEMSE}
\end{figure}

The Fig. \ref{fig:sim3-CEMSE} demonstrates the convergence of the constructed estimates $\va_{j,t}$'s to the parameter of interest $\vwo$ in the mean-square sense. We point out that the recursions \eqref{eq:equation2} and \eqref{eq:equation3} also provide the global mean-square deviation of the constructed estimates for the certain combination weight $\mSig$ in Table \ref{tab:sim2} and the theoretical recursion matches with the simulated results.

\begin{figure}[t!]
\centering
\includegraphics[width = 0.35\textwidth]{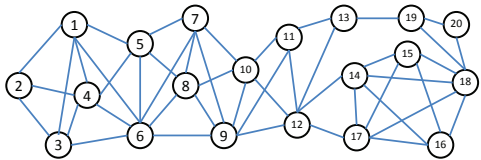}
\caption{Network topology with $N = 20$ for the Example 2.}
\label{fig:scenario}
\end{figure}

\begin{figure}[t!]
\centering
\includegraphics[width = 0.4\textwidth]{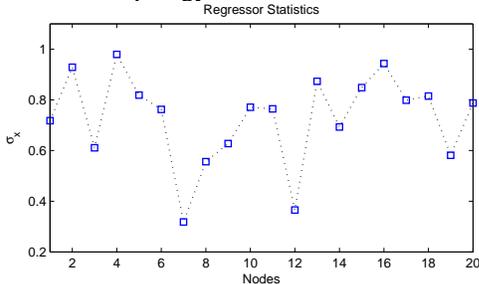}
\caption{Network statistical profile ($\sigma_{n_i} = 0.1$).}
\label{fig:reg}
\end{figure}

\begin{figure}[t!]
\centering
\includegraphics[width = 0.4\textwidth]{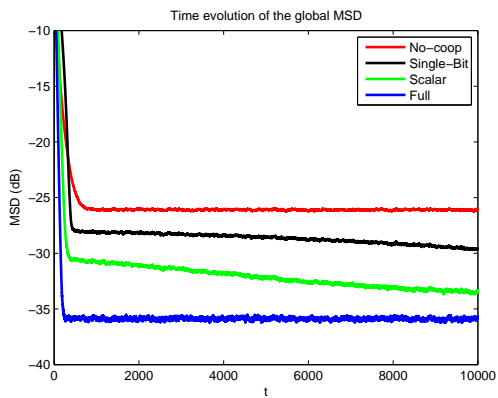}
\caption{The global EMSE curves in relatively large network size and long filter length and the confidence parameter is adaptively chosen for single-bit and scalar diffusion strategies.}
\label{fig:sim4}
\end{figure}

In the second example, we examine the convergence performance of the adaptive confidence parameter in relatively large network $N = 20$ with long filter length $M = 10$ (See Fig. \ref{fig:scenario}). We again observe a stationary data $d_{i,t} = \vu_{i,t}^T\vwo + v_{i,t}$ for $i \in \{1,2,\cdots,N\}$. The regressor data $\vu_{i,t}$ is zero-mean i.i.d. Gaussian whose standard deviation varies over the network as in Fig. \ref{fig:reg}. The observation noise $v_{i,t}$ is zero-mean i.i.d. Gaussian whose variance is $\sigma_{n_i} = 10^{-2}$. We note that the signal-to-noise ratio varies from $10$ to $100$ over the network similar to the example $1$. The standard deviation of the projection operator $\vec{c}_{i,t}$ is $\sigma_{c_i} = 1$ and the parameter of interest $\vwo \in \mbox{$\mathbbm{R}^{10}$}$ is randomly chosen.

We again use the Metropolis rule as the combination rule, however, in this example, we adapt the confidence parameter through \eqref{eq:sigmoid} and \eqref{eq:update} where we resort to the convex mixture of the adaptive filtering algorithms \cite{garcia2005,garcia2006,silva2008,kozat}. We also choose the step sizes the same for the distributed LMS update \eqref{eq:dist} of all configurations at all nodes, i.e., $\mu_i = 0.042$. In example 2, the step sizes for the construction update \eqref{eq:construction} are $\eta_i = 0.0042$ (for single-bit approach) and $\eta_i = 0.1$ (for one-dimension diffusion approach). We set $\mu_{\mathrm{cvx}} = 10$ in \eqref{eq:update}. The Fig. \ref{fig:sim4} shows the global MSD curves of the no-cooperation, single-bit, scalar and full diffusion strategies. We observe that the adaptive confidence parameter improves the convergence performance of the compressive diffusion strategies far more such that they achieve comparable performance while the reduction of the communication load is tremendous.
\vspace{-0.2in}
\section{Conclusion}
In the diffusion based distributed estimation strategies, the communication load increases far more in the large networks or highly connected network of nodes. Hence, the compressive diffusion approach plays an essential role in achieving comparable convergence performance with the full diffusion configurations while reducing the communication load significantly. We provide a complete performance analysis for the compressive diffusion strategies. We analyze the mean-square convergence, steady-state behavior and the tracking performance of the scalar and single-bit diffusion approaches. The numerical examples show the theoretical analysis model the simulated results accurately. Additionally, we introduce the confidence parameter concept, which adds one more freedom of dimension to the combination rule in order to improve the convergence performance. When we adapt the confidence parameter using the well-known mixture algorithms, we observe enormous enhancement in the convergence performance of the compressive diffusion strategies even for the relatively long filter lengths.
\vspace{-0.2in}
\appendices
\section{Proof for Lemma 1}
We first show the equality of \eqref{eq:lem1} for the two-node case. Then the extension for a larger network is straight forward.
We can rewrite the term on the left hand side (LHS) of \eqref{eq:lem1} as
\begin{align}\label{ap:1}
E[\vtpsi_t^T\cXu^T\mSig_2\mN\vC_t\sign(\veps_t)]& \nn\\
&\hspace{-2.3cm}=E\left[\vtpsi_t^T\cXu^T
\underbrace{\begin{bmatrix} \vect{\varsigma_1} & \vect{\varsigma_2} \\ \vect{\varsigma_3} & \vect{\varsigma_4} \end{bmatrix}}_{\mSig_2}
\mN\vC_t\sign(\veps_t)\right].
\end{align}
After some algebra, \eqref{ap:1} yields
\begin{align}\label{ap:2}
E[\vtpsi_t^T\cXu^T\mSig_2\mN\vC_t\sign(\veps_t)]& \nn\\
&\hspace{-2.8cm}= E[(\gamma_{11}\vtphi_{1,t}^T + \gamma_{12}\vta_{2,t}^T)\vect{\varsigma_1}\eta_1\vc_{1,t}\sign(\eps_{1,t})]\nn\\
&\hspace{-2.4cm}+ E[(\gamma_{11}\vtphi_{1,t}^T + \gamma_{12}\vta_{2,t}^T)\vect{\varsigma_2}\eta_2\vc_{2,t}\sign(\eps_{2,t})]\nn\\
&\hspace{-2.4cm}+ E[(\gamma_{22}\vtphi_{2,t}^T + \gamma_{21}\vta_{1,t}^T)\vect{\varsigma_3}\eta_1\vc_{1,t}\sign(\eps_{1,t})]\nn\\
&\hspace{-2.4cm}+ E[(\gamma_{22}\vtphi_{2,t}^T + \gamma_{21}\vta_{1,t}^T)\vect{\varsigma_4}\eta_2\vc_{2,t}\sign(\eps_{2,t})].
\end{align}

In order to evaluate the expectations on the RHS of \eqref{ap:2}, we assume that the step sizes are sufficiently small and filter is sufficiently long so that the deviation terms changes negligibly slow with respect to the regressor data $\vc_{i,t}$. Then, according to the Price's result \cite{price1958,mcmahon1964}, we obtain
\begin{align}\label{ap:3}
E[\vtpsi_t^T\cXu^T\mSig_2\mN\vC_t\sign(\veps_t)]& \nn\\
&\hspace{-3cm}= E[(\gamma_{11}\vtphi_{1,t}^T + \gamma_{12}\vta_{2,t}^T)\vect{\varsigma_1}\eta_1\vc_{1,t}\eps_{1,t}]
\frac{E|\eps_{1,t}|}{E[\eps_{1,t}^2]}\nn\\
&\hspace{-2.7cm}+ E[(\gamma_{11}\vtphi_{1,t}^T + \gamma_{12}\vta_{2,t}^T)\vect{\varsigma_2}\eta_2\vc_{2,t}\eps_{2,t}]
\frac{E|\eps_{2,t}|}{E[\eps_{2,t}^2]}\nn\\
&\hspace{-2.7cm}+ E[(\gamma_{22}\vtphi_{2,t}^T + \gamma_{21}\vta_{1,t}^T)\vect{\varsigma_3}\eta_1\vc_{1,t}\eps_{1,t}]
\frac{E|\eps_{1,t}|}{E[\eps_{1,t}^2]}\nn\\
&\hspace{-2.7cm}+ E[(\gamma_{22}\vtphi_{2,t}^T + \gamma_{21}\vta_{1,t}^T)\vect{\varsigma_4}\eta_2\vc_{2,t}\eps_{2,t}]
\frac{E|\eps_{2,t}|}{E[\eps_{2,t}^2]}.
\end{align}
Rearranging \eqref{ap:3} into a matrix product form leads \eqref{eq:lem1}. Following the same way, we can also get \eqref{eq:lem2} and the proof is concluded.
\vspace{-0.2in}
\section{Proof for Lemma 2}
We derive the RHS of \eqref{eq:lem3} for the two-node case for simplicity, however, it also satisfies any order of network. For two-node case, the LHS of \eqref{eq:lem3} yields
\begin{align}
E\left[\sign(\veps_t)^T\vC_t^T\mN\mSig_4\mN\vC_t\sign(\veps_t)\right]&\nn\\
&\hspace{-3.8cm}=E\left[\sign(\eps_{1,t})\vc_{1,t}^T\eta_1\vect{\varsigma_1}\eta_1\vc_{1,t}\sign(\eps_{1,t})\right]\nn\\
&\hspace{-3.3cm}+E\left[\sign(\eps_{1,t})\vc_{1,t}^T\eta_1\vect{\varsigma_2}\eta_2\vc_{2,t}\sign(\eps_{2,t})\right]\nn\\
&\hspace{-3.3cm}+E\left[\sign(\eps_{2,t})\vc_{2,t}^T\eta_2\vect{\varsigma_3}\eta_1\vc_{1,t}\sign(\eps_{1,t})\right]\nn\\
&\hspace{-3.3cm}+E\left[\sign(\eps_{2,t})\vc_{2,t}^T\eta_2\vect{\varsigma_4}\eta_2\vc_{2,t}\sign(\eps_{2,t})\right].\nn
\end{align}
We re-emphasize that the regressor $\vc_{i,t}$ is spatially and temporarily independent. Hence, we obtain
\begin{align}
E\left[\sign(\veps_t)^T\vC_t^T\mN\mSig_4\mN\vC_t\sign(\veps_t)\right]&\nn\\
&\hspace{-4.2cm}=E\left[\vc_{1,t}^T\eta_1\vect{\varsigma_1}\eta_1\vc_{1,t}\right]+E\left[\vc_{2,t}^T\eta_2\vect{\varsigma_4}\eta_2\vc_{2,t}\right]\nn\\
&\hspace{-4cm}+E\left[\vc_{1,t}\sign(\eps_{1,t})\right]^T\eta_1\vect{\varsigma_2}\eta_2E\left[\vc_{2,t}\sign(\eps_{2,t})\right]\nn\\
&\hspace{-4cm}+E\left[\vc_{2,t}\sign(\eps_{2,t})\right]^T\eta_2\vect{\varsigma_3}\eta_1E\left[\vc_{1,t}\sign(\eps_{1,t})\right].\label{ap:4}
\end{align}
Using Price's result, we can evaluate the last two terms on the RHS of \eqref{ap:4} as follows
\begin{equation}
E\left[\vc_{1,t}\sign(\eps_{1,t})\right] = \frac{E|\eps_{1,t}|}{E[\eps_{1,t}^2]}E\left[\vc_{1,t}\eps_{1,t}\right]\nn
\end{equation}
and
\begin{equation}
E\left[\vc_{2,t}\sign(\eps_{2,t})\right] = \frac{E|\eps_{2,t}|}{E[\eps_{2,t}^2]}E\left[\vc_{2,t}\eps_{2,t}\right].\nn
\end{equation}
We point out that the terms involving the diagonal entries of the weighting matrix $\mSig_4$ in \eqref{ap:4} does not include the deviation terms. As a result, rearranging \eqref{ap:4} into a compact form results in \eqref{eq:lem3}. This concludes the proof.

{\def\ninept{\def\baselinestretch{0.8}}
\ninept
\bibliographystyle{IEEEtran}
\bibliography{my_references}}

\end{document}